\documentclass{article}

\PassOptionsToPackage{numbers,sort&compress}{natbib}
\usepackage[final]{neurips_2021}

\usepackage{filecontents}
\usepackage[utf8]{inputenc} 
\usepackage[T1]{fontenc}    
\usepackage{url}            
\usepackage{booktabs}       
\usepackage{nicefrac}       
\usepackage{xcolor}         
\usepackage{graphicx} 
\usepackage[hidelinks]{hyperref}       
\usepackage{tabularx}
\usepackage{amsmath,amsfonts,amssymb} \usepackage{bbold}       
\usepackage{bm}
\usepackage{dsfont}
\usepackage[final]{microtype}      
\usepackage{multirow} 
\colorlet{check}{red}

\usepackage{xr}
\makeatletter
\newcommand*{\addFileDependency}[1]{
  \typeout{(#1)}
  \@addtofilelist{#1}
  \IfFileExists{#1}{}{\typeout{No file #1.}}
}
\makeatother

\newcommand*{\myexternaldocument}[1]{
    \externaldocument[S]{#1}
    \addFileDependency{#1.tex}
    \addFileDependency{#1.aux}
}

\myexternaldocument{supplement}

\title{SE(3)-equivariant prediction of molecular wavefunctions and electronic densities}

\author{%
  Oliver T. Unke$^{1,2,8,10}$\thanks{These authors contributed equally.}\enspace\thanks{Work done at TU Berlin prior to joining Google Research.} \\
  \texttt{oliver.unke@gmail.com} \\
  \And
  Mihail Bogojeski$^{1,8*}$ \\
  \texttt{mihailbogojeski@gmail.com} \\
  \AND
  Michael Gastegger$^{1,2,9}$ \\
  \And
  Mario Geiger$^{3}$ \\
  \And
  Tess Smidt$^{4,5}$ \\
  \AND
  Klaus-Robert M\"uller $^{1,6,7,8,10}$ \\
  \texttt{klaus-robert.mueller@tu-berlin.de}\\
  \\
  \footnotesize
  $^1$ Machine Learning Group, Technische Universit\"at Berlin,
  10587 Berlin, Germany \\
  \footnotesize
  $^2$ DFG Cluster of Excellence ``Unifying Systems in Catalysis'' (UniSysCat), Technische Universit\"at Berlin,\\
  \footnotesize
  10623 Berlin, Germany \\
  \footnotesize
  $^3$ Institute of Physics, \'Ecole Polytechnique F\'ed\'erale de Lausanne,  1015 Lausanne, Switzerland \\
  \footnotesize
  $^4$ Computational Research Division, Lawrence Berkeley National Laboratory, Berkeley, CA 94720\\
  \footnotesize
  $^5$ Center for Advanced Mathematics for Energy Research Applications (CAMERA),\\
  \footnotesize
  Lawrence Berkeley National Laboratory, Berkeley, CA 94720\\
  \footnotesize
  $^6$ Department of Artificial Intelligence, Korea University, Anam-dong, Seongbuk-gu, Seoul 02841, Korea \\
  \footnotesize
  $^7$ Max Planck Institute for Informatics, Stuhlsatzenhausweg, 66123 Saarbr\"ucken, Germany \\
  \footnotesize $^8$ BIFOLD -- Berlin Institute for the Foundations of Learning and Data,  Berlin, Germany \\
  \footnotesize
  $^9$ BASLEARN -- TU Berlin/BASF Joint Lab for Machine Learning, Technische Universit{\"a}t Berlin,\\
  \footnotesize
  10587 Berlin, Germany\\
  \footnotesize $^{10}$ Google Research, Brain Team, Berlin, Germany \\
}

\begin{document}

\maketitle

\begin{abstract}
Machine learning has enabled the prediction of quantum chemical properties with high accuracy and efficiency, allowing to bypass computationally costly \textit{ab initio} calculations. Instead of training on a fixed set of properties, more recent approaches attempt to learn the electronic wavefunction (or density) as a central quantity of atomistic systems, from which all other observables can be derived. This is complicated by the fact that wavefunctions transform non-trivially under molecular rotations, which makes them a challenging prediction target. To solve this issue, we introduce general SE(3)-equivariant operations and building blocks for constructing deep learning architectures for geometric point cloud data and apply them to reconstruct wavefunctions of atomistic systems with unprecedented accuracy.
Our model achieves speedups of over three orders of magnitude compared to \textit{ab initio} methods and reduces prediction errors by up to two orders of magnitude compared to the previous state-of-the-art. This accuracy makes it possible to derive properties such as energies and forces directly from the wavefunction in an end-to-end manner. We demonstrate the potential of our approach in a transfer learning application, where a model trained on low accuracy reference wavefunctions implicitly learns to correct for electronic many-body interactions from observables computed at a higher level of theory. Such machine-learned wavefunction surrogates pave the way towards novel semi-empirical methods, offering resolution at an electronic level while drastically decreasing computational cost. Additionally, the predicted wavefunctions can serve as initial guess in conventional \textit{ab initio} methods, decreasing the number of iterations required to arrive at a converged solution, thus leading to significant speedups without any loss of accuracy or robustness. While we focus on physics applications in this contribution, the proposed equivariant framework for deep learning on point clouds is promising also beyond, say, in computer vision or graphics. 
\end{abstract}

\section{Introduction}
\label{sec:introduction}
Machine learning (ML) methods are becoming increasingly popular in quantum chemistry as a means to circumvent expensive \textit{ab initio} calculations, and led to advances in a broad range of applications, including the construction of potential energy surfaces~\citep{behler2007generalized, bartok2010gaussian, smith2017ani, chmiela2017machine, chmiela2018towards, christensen2019operators, Batzner2021_e3nn_molecular_dynamics,unke2020machine}, prediction of electron densities and density functionals~\citep{snyder2012finding,brockherde2017bypassing, ryczko2018deep,grisafi2018transferable,nagai2020completing, bogojeski2020quantum}, and development of models capable of predicting a range of physical observables across chemical space~\citep{rupp2012fast,Montavon2013a,de2016comparing, schutt2017quantum,schutt2017schnet,schutt2018schnet,pronobis2018many, eickenberg2018solid,unke2019physnet,von2020exploring,Chen2021_e3nn_phonon, schutt2021equivariant, unke2021spookynet,keith2021combining,gastegger2021machine}. Typically, such models are trained on reference data for a predetermined set of quantum chemical properties and need to be retrained if other properties are required. However, if a model is capable of predicting the wavefunction, expectation values for \emph{any} observable can be derived from it. Unfortunately, such an approach is complicated by the fact that wavefunctions are typically expressed in terms of rotationally equivariant basis functions, introducing non-trivial transformations under molecular rotations, which are difficult to learn from data. To solve this issue, we propose several SE(3)-equivariant operations for deep learning architectures for geometric point cloud data, which capture the effects of translations and rotations without needing to learn them explicitly. We assemble these building blocks to construct PhiSNet, a novel 
deep learning (DL) architecture for predicting wavefunctions and electronic densities, which is significantly more accurate than non-equivariant models. For the first time, sufficient accuracy is reached to predict properties like energies and forces directly from the wavefunction  and in  end-to-end manner.

This makes it possible to learn wavefunctions that lead to modified properties, which is interesting from an inverse design perspective; or the development of novel machine-learned semi-empirical methods, for example by learning a correction to the wavefunction that mimics the effects of electron correlation. Such hybrid methods maintain the accuracy and generality of high level electronic structure calculations while drastically reducing their computational cost. In addition, the predicted wavefunctions can serve as initial guess to speed up conventional \textit{ab initio} methods.\\ 
Beyond physics, other applications of our proposed equivariant DL architecture to e.g.\ computer vision or graphics are conceivable -- whenever accurate invariant analyses of high dimensional point clouds are of importance.  

In summary, this work provides the following contributions:
\begin{itemize}
  \item We describe general SE(3)-equivariant operations and building blocks for constructing DL architectures for geometric point cloud data.
  \item We propose PhiSNet, a neural network for predicting wavefunctions and electronic densities from equivariant atomic representations, ensuring physically correct transformation under translations and rotations.
  \item We apply PhiSNet to predict wavefunctions and electronic densities of several molecules and show that our model reduces prediction errors of electronic structure properties by a factor of up to two orders of magnitude compared to the previous state-of-the-art and achieves speedups of over three orders of magnitude compared to \textit{ab initio} solutions.
  \item We showcase a novel transfer-learning application, where a model trained on low accuracy wavefunctions is adapted to predict properties computed at a higher level of theory by learning a correction that implicitly captures the effects of many-body electron correlation.
  \item We demonstrate that the predicted wavefunctions can serve as initial guess in conventional quantum chemistry methods, leading to significant speedups without sacrificing the accuracy or robustness of \textit{ab initio} solutions.
\end{itemize}

In principle, our method could also be used to construct orbital features as inputs for methods like OrbNet~\cite{qiao2020orbnet}, which otherwise rely on semi-empirical or \textit{ab initio} methods.

\section{Related work}
\label{sec:related_work}
Only a small number of studies apply ML to the challenging problem of modeling the wavefunction directly \cite{carleo2017solving}. This is usually done by predicting Hamiltonian matrices, from which the wavefunction can be obtained by solving a generalized eigenvalue problem. The earliest such study we are aware of is by \citet{hegde2017machine}, who used kernel ridge regression to learn the Hamiltonian matrix for two simple case studies. Later, \citet{schutt2019unifying} proposed the SchNOrb neural network architecture, which constructs the Hamiltonian matrix of molecules in a block-wise manner from atom-pair features. Recently, \citet{li2021deep} presented a deep neural network architecture for predicting the Hamiltonian matrix of simple periodic crystals. However, none of these models form their predictions in a rotationally equivariant manner, i.e.\ they need to learn how to predict the Hamiltonian matrix for all possible orientations of the system of interest, which requires large amounts of training data. Even when data augmentation via random rotations~\citep{montavon2012learning} or special Hamiltonian representations~\citep{gastegger2020deep} are used to mitigate this issue, the final model is only approximately equivariant, i.e.\ properties derived from the wavefunction can change unphysically when the system is rotated or the frame of reference is changed. Here, we draw upon insights from a range of SE(3)-equivariant models~\citep{cohen2016group,marcos2017rotation,cohen2018spherical,kondor2018covariant,weiler20183d,thomas2018tensor,Grisafi2018,hy2018predicting,NEURIPS2019_b9cfe8b6,NEURIPS2019_ea9268cb,cohen2019gauge,fuchs2020se} to ensure that predictions exactly preserve the physically correct dependence with respect to the orientation of inputs (for a more detailed discussion see the supplement). We would like to remark that after the initial submission of this manuscript, \citet{nigam2021equivariant} introduced a complementary method to construct equivariant representations for Hamiltonian matrices, e.g.\ for the use in kernel machines.

\section{Background}
\label{sec:background}

\begin{figure}
  \centering
	\includegraphics[width=\textwidth]{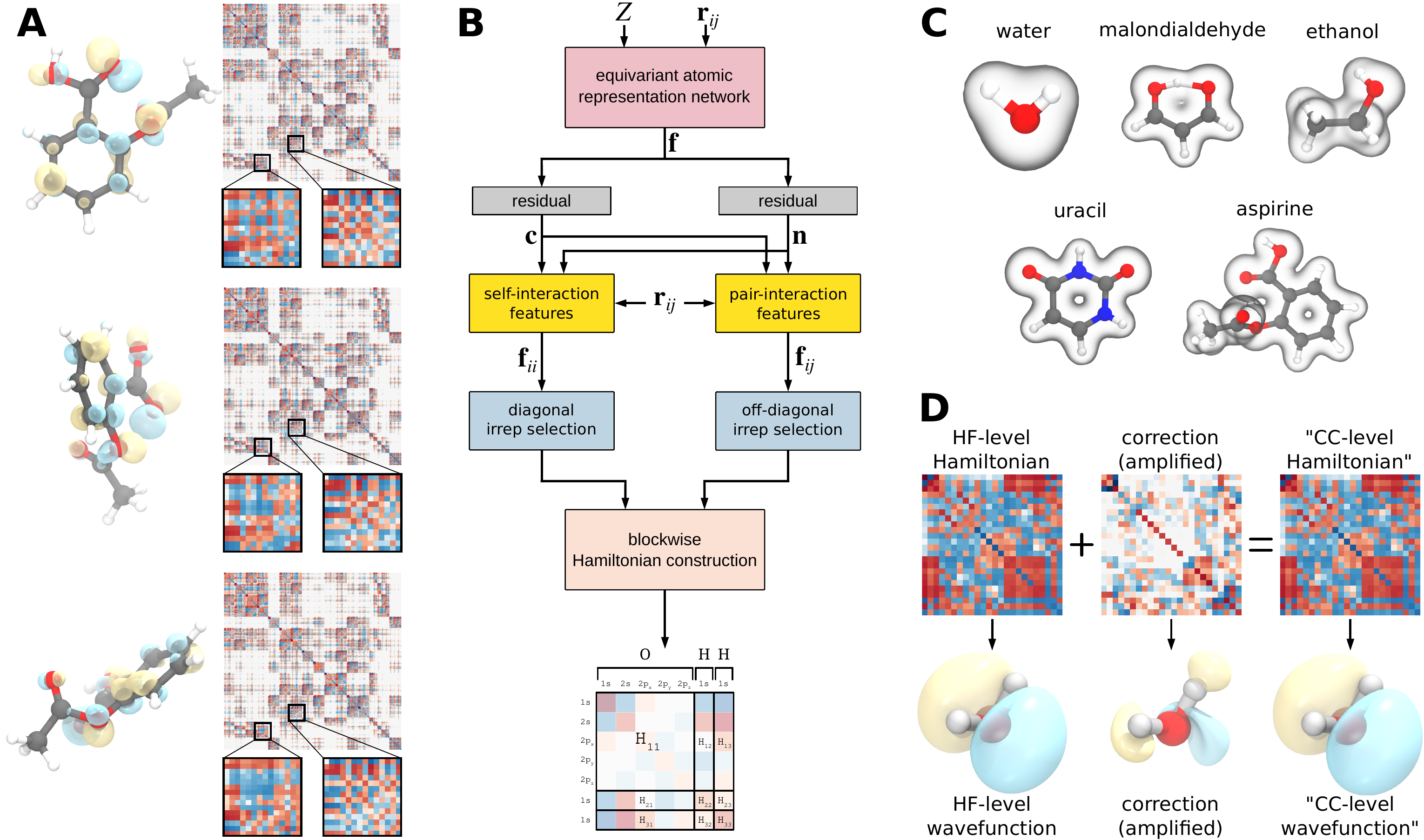}
  \caption{\footnotesize \textbf{A}: Illustration of an aspirine molecule and its highest occupied molecular orbital (HOMO) in three different orientations, showing how the wavefunction (left) and Hamiltonian matrix (right) change with respect to rotations.  \textbf{B}: Overview of the proposed PhiSNet architecture. The atomic representation network creates atom-wise equivariant features, which are used to produce self-interaction and pair-interaction features (Fig.~\ref{Sfig:nn_architecture}), from which the Hamiltonian matrix is constructed block-by-block (Fig.~\ref{Sfig:hamiltonian_composition}). \textbf{C}: Visualisation of electronic densities (squared wavefunction) of various molecules predicted with our approach. \textbf{D}: Illustration of a transfer learning application, where
  a model pretrained on Hartree-Fock (HF) Hamiltonians is fine-tuned to match energies and forces derived from highly accurate coupled cluster (CC) calculations. The model achieves this by learning a correction to the Hamiltonian matrix, which mimics the effects of many-body electron correlation. The effective ``CC-level'' Hamiltonian can be thought of as a HF-level Hamiltonian plus a correction term. The HOMO is shown to visualize subtle changes to the wavefunction (the correction is amplified in magnitude by a factor of $10^3$ for better visibility).}
	\label{fig:overview}
\end{figure}

The aim of most quantum chemistry methods is to solve the electronic Schrödinger equation
\begin{equation}
\hat{H}_\mathrm{el} \Psi_\mathrm{el} = E_\mathrm{el} \Psi_\mathrm{el},
\label{eq:schrodinger_equation}
\end{equation}
where $\hat{H}_\mathrm{el}$ is the Hamiltonian operator describing the interactions and motion of the electrons, $\Psi_\mathrm{el}$ is the electronic wavefunction and $E_\mathrm{el}$ is the ground state energy.
After $\Psi_\mathrm{el}$ is determined, all physical observables (beyond $E_\mathrm{el}$) can be derived by applying other operators (similar to $\hat{H}_\mathrm{el}$) to the wavefunction and reading out the corresponding eigenvalues~\citep{schechter2012operator}. In practice, Eq.~\ref{eq:schrodinger_equation} is usually solved by expressing $\Psi_\mathrm{el}$ as an antisymmetric product of molecular orbitals 
\begin{equation}
\psi_i = \sum_j C_{ij} \phi_j\,,
\label{eq:molecular_orbitals}
\end{equation}
which are written as linear combinations of atom-centered basis functions $\phi$. This leads to the equation
\begin{equation}
\mathbf{H}\mathbf{C} = \bm{\epsilon}\mathbf{S}\mathbf{C}\,,
\label{eq:schrodinger_matrix_equation}
\end{equation}
where the Hamiltonian is written as a matrix $\mathbf{H}$ with entries $H_{ij} = \int \phi_i^*(\mathbf{r}) \hat{H}_{\mathrm{el}} \phi_j(\mathbf{r}) d\mathbf{r}$  ($\mathbf{r}$ denotes the electronic coordinates). The overlap matrix $\mathbf{S}$ with entries $S_{ij} = \int \phi_i^*(\mathbf{r})\phi_j(\mathbf{r}) d\mathbf{r}$ has to be introduced and a generalized eigenvalue problem must be solved, because the basis functions $\phi$ are typically not orthonormal. The eigenvectors $\mathbf{C}$ specify the wavefunction $\Psi_\mathrm{el}$ via the coefficients $C_{ij}$ of the molecular orbitals (Eq.~\ref{eq:molecular_orbitals}) and the eigenvalues $\bm{\epsilon}$ are the corresponding orbital energies. Since eigenvectors are only defined up to sign changes, predicting $\mathbf{H}$ (instead of $\mathbf{C}$) is preferable for ML applications. This is a challenging task, because the basis functions are typically products of a radial component and spherical harmonics, which introduces non-trivial dependencies of the matrix elements with respect to the orientation of the chemical system (see Fig.~\ref{fig:overview}A). Spherical harmonics $Y^{m}_{l}$ of degree $l=0,\dots,\infty$ and order $m=-l,\dots,l$ form a complete orthonormal basis for functions on the surface of a sphere and can be used to derive irreducible representations (irreps) of the 3D rotation group $\mathrm{SO}(3)$. In other words, they are rotationally equivariant, which means that when $\mathbf{r}$ is rotated, the values of $Y^{m}_{l}(\mathbf{r})$ change accordingly. Since $\Psi_\mathrm{el}$ is expressed with spherical harmonic basis functions, the entries of $\mathbf{H}$ transform predictably under rotations (a more detailed overview of quantum chemistry fundamentals, groups, equivariance, and spherical harmonics is given in Section~\ref{Ssec:detailed_background} of the supplement).

\section{Deep learning architecture for molecular wavefunctions}
Deep message-passing neural networks (MPNNs) \citep{gilmer2017neural} for quantum chemistry applications, such as DTNN~\citep{schutt2017quantum} or SchNet~\citep{schutt2017schnet}, model physical properties of chemical systems as a sum over atomic contributions predicted from features $\mathbf{x}_i \in \mathbb{R}^F$ for each atom~$i$. Starting from initial element-specific embeddings, the features are constructed by iteratively exchanging ``messages'' between neighboring atoms $i$ and $j$, which depend on the current $\mathbf{x}_i$ and $\mathbf{x}_j$ and their distance $r_{ij}$. Since geometric information enters only in the form of pairwise distances, the final atomic features are rotationally invariant by construction. This is desirable when they are used to predict a quantity that itself is rotationally invariant, for example the potential energy. To predict observables that change under rotation, e.g.\ electric moments or the electronic Hamiltonian, a natural extension is to instead construct rotationally equivariant features. To see how this can be achieved, it is useful to think of the $F$ entries of atomic feature vectors $\mathbf{x}_i\in \mathbb{R}^F$ as different ``channels'', where each channel carries scalar information about the chemical environment of atom~$i$. To construct rotationally equivariant features, each scalar channel can be replaced by values derived from the spherical harmonics up to a maximum degree $L$, i.e.\ there are now $F\times(L+1)^2$ entries (each spherical harmonic degree $l \in \{0,\dots,L\}$ contributes $2l+1$ values for all possible orders $m\in \{-l,\dots,l\}$). Our proposed model, which we call PhiSNet, shares basic design principles with PhysNet~\citep{unke2019physnet}, but uses equivariant (instead of invariant) operations throughout its architecture. Contrary to most other MPNNs, instead of directly predicting chemical properties from atomic features, PhiSNet constructs the Hamiltonian matrix in a block-wise manner from equivariant representations. All known physical symmetries of the Hamiltonian are preserved by construction, an essential constraint which is not satisfied by existing DL architectures for chemical applications. 

\paragraph{Notation} Whenever equivariant features are discussed, bold symbols (e.g.\ $\mathbf{x}$) refer to the collection of all $F\times(L+1)^2$ entries for all feature channels $F$ and spherical harmonics degrees $l\in\{0,\dots,L\}$, whereas a superscript $l$ in parentheses (e.g.\ $\mathbf{x}^{(l)}$) is used to refer only to the  $F\times (2l+1)$ entries of degree $l$. Similarly,  $\mathbf{Y}(\mathbf{r})$ refers to the collection of all $1\times (L+1)^2$ spherical harmonics with distinct combinations of $l$ and $m$, whereas $\mathbf{Y}^{(l)}(\mathbf{r})$ refers to the $1\times (2l+1)$ values for degree $l$. The notation $\mathbf{c}\circ\mathbf{x}$ denotes a Hadamard product between matrices, i.e.\ $\mathbf{c}\in \mathbb{R}^{F\times (L+1)^2}$ and $\mathbf{x}\in \mathbb{R}^{F\times (L+1)^2}$ are multiplied entry-wise. When there is no one-to-one correspondence between the entries in $\mathbf{c}$ and $\mathbf{x}$, e.g.\ when $\mathbf{c}\in \mathbb{R}^{F}$ and $\mathbf{x}\in \mathbb{R}^{F\times (L+1)^2}$, $\mathbf{c}\circ\mathbf{x}$ implies that $\mathbf{c}$ is ``broadcasted'' across the missing dimensions, i.e.\ each of the $(L+1)^2$ ``slices'' of $\mathbf{x}$ is multiplied entry-wise with $\mathbf{c}$ and the result $\mathbf{c}\circ\mathbf{x}$ has dimensions $F\times (L+1)^2$.
Double-struck digits denote an irreducible representation (irrep) with the corresponding number of dimensions $2l+1$. For example, $\mathbb{1}$ refers to a one-dimensional irrep of degree $l=0$ and $\mathbb{3}$ to a three-dimensional irrep of degree $l=1$. By abuse of terminology, the term ``irrep'' is also used for the individual $(2l+1)$-dimensional components along the $F$ feature dimensions of $\mathbf{x} \in \mathbb{R}^{F\times(L+1)^2}$. When a collection of equivariant features $\mathbf{x}^{(l)} \in \mathbb{R}^{F_{\rm in}\times(2l+1)}$ is multiplied by a matrix $\mathbf{M} \in \mathbb{R}^{F_{\rm out}\times F_{\rm in}}$, the result is $\mathbf{M}\mathbf{x}^{(l)} \in \mathbb{R}^{F_{\rm out}\times(2l+1)}$, i.e.\ the ordinary rules for matrix multiplication apply.

\subsection{SE(3)-equivariant neural network building blocks}
\label{subsec:building_blocks}

In the following, we describe general-purpose operations for building SE(3)-equivariant MPNNs, which can also be used outside a chemical context to build feature representations for other point cloud data. Additionally, we discuss necessary modifications to established neural network components (e.g.\ linear layers or activation functions) for keeping feature representations rotationally equivariant.

\paragraph{Activation functions} may only be applied to scalar ($l=0$) features, or else the output loses its equivariant properties:
\begin{equation}
\boldsymbol{\sigma}(\mathbf{x})^{(l)} =
\begin{cases}
\sigma(\mathbf{x}^{(l)}) & l = 0 \\
\mathbf{x}^{(l)} & l > 0
\end{cases}\,.
\label{eq:activation_function1}
\end{equation}
Here, $\sigma$ can be any activation function and the notation $\sigma(\mathbf{x}^{(l)})$ means that $\sigma$ is applied to $\mathbf{x}^{(l)}$ entry-wise. In this work, a generalized SiLU~\citep{hendrycks2016gaussian,elfwing2018sigmoid}  activation function (also known as Swish~\citep{ramachandran2017searching}) given by
\begin{equation}
\sigma(x) =  \frac{\alpha x}{1+e^{-\beta x}} 
\label{eq:activation_function2}
\end{equation}
is used, where $\alpha$ and $\beta$ are both learned and separate parameters are kept for all feature channels and instances of $\sigma$ \citep{unke2021spookynet} (see Section~\ref{Ssubsec:activation_function} in the supplement for additional details).

\paragraph{Linear layers} are applied to each degree $l$ according to
\begin{equation}
\mathrm{linear}_{F_{\rm in} \to F_{\rm out}}(\mathbf{x})^{(l)} =
\begin{cases}
\mathbf{W}_l\mathbf{x}^{(l)} + \mathbf{b} & l = 0 \\
\mathbf{W}_l\mathbf{x}^{(l)} & l > 0
\end{cases}\,,
\label{eq:linear_layer}
\end{equation}
where $\mathbf{W} \in \mathbb{R}^{F_{\rm out}\times F_{\rm in}}$ and $\mathbf{b} \in \mathbb{R}^{F_{\rm out}}$ are weights and biases, respectively. The subscript $l$ is used to distinguish the weights for different degrees $l$, i.e.\ separate linear transformations are applied to the features of each degree $l$. The bias term must be omitted for $l>0$ so that output features stay rotationally equivariant.

\paragraph{Tensor product contractions} are used to couple two equivariant feature representations $\mathbf{x}^{(l_1)}$ and $\mathbf{y}^{(l_2)}$ to form new features $\mathbf{z}^{(l_3)}$. The (reducible) tensor product $\mathbf{x}^{(l_1)}\otimes \mathbf{y}^{(l_2)}$ of two irreps has $(2l_1+1)(2l_2+1)$ dimensions and can be expanded into a direct sum of irreducible representations, e.g.\ $\mathbb{3}\otimes \mathbb{5} = \mathbb{3} \oplus \mathbb{5} \oplus \mathbb{7}$. In general, the value for order $m_3$ of the irrep of degree $l_3\in \{|l_1-l_2|,\dots ,l_1+l_2\}$  in the direct sum representation of the tensor product $\mathbf{x}^{(l_1)}\otimes \mathbf{y}^{(l_2)}$ is given by
\begin{equation}
  \left(\mathbf{x}^{(l_1)}\otimes \mathbf{y}^{(l_2)}\right)^{l_3}_{m_3} = \sum_{m_1=-l_1}^{l_1} \sum_{m_2=-l_2}^{l_2} C^{l_3,l_2,l_1}_{m_3,m_2,m_1} x^{l_1}_{m_1} y^{l_2}_{m_2}\,,
\label{eq:tensor_product_contraction}
\end{equation}
where $C^{l_3,l_2,l_1}_{m_3,m_2,m_1}$ are Clebsch-Gordan coefficients (CGCs)~\citep{varshalovich1988quantum}. The short-hand notation $\mathbf{x}^{(l_1)}\underset{l_3}{\otimes} \mathbf{y}^{(l_2)}$ is used to refer to the irrep of degree $l_3$ in the direct sum representation of $\mathbf{x}^{(l_1)}\otimes \mathbf{y}^{(l_2)}$. In other words, the operation $\mathbf{x}^{(l_1)}\underset{l_3}{\otimes} \mathbf{y}^{(l_2)}$ performs the tensor product and contracts the result to a single irrep of degree $l_3$. A similar construction is also used in Clebsh-Gordan~\cite{kondor2018clebsch} and Tensor Field networks~\cite{thomas2018tensor}.

\paragraph{Tensor product expansions} are inverse to tensor product contractions. Instead of contracting two irreps into one, CGCs are used to expand a single irrep $\mathbf{x}^{(l_3)}$ into a $(2l_1 + 1)\times (2l_2 + 1)$ matrix that represents its contribution to the direct sum representation of the tensor product of two irreps of degree $l_1$ and $l_2$, where $|l_2 - l_1| \leq l_3 \leq l_2 + l_1$:

\begin{equation}
  \left(\overline{\otimes}\mathbf{x}^{(l_3)}\right)^{l_1, l_2}_{m_1, m_2}  = \sum_{m_3=-l_3}^{l_3} \mathbf{C}^{l_1,l_2,l_3}_{m_l,m_2,m_3} x^{l_3}_{m_3}\,.
\label{eq:tensor_product_expansion}
\end{equation}
Mirroring the shorthand used for tensor product contractions, $\overset{l_1,l_2}{\overline{\otimes}}\mathbf{x}^{(l_3)}$ will be used to refer to the $(2l_1 + 1)\times (2l_2 + 1)$ matrix that is obtained from the tensor product expansion of $\mathbf{x}^{(l_3)}$. 

\paragraph{Selfmix layers} are used to recombine (``mix'') the $F$ features of a single input $\mathbf{x}\in \mathbb{R}^{F\times (L_{\rm in}+1)^2}$ across different degrees and optionally allow changing the maximum degree from $L_{\rm in}$ to $L_{\rm out}$. The output features of degree $l_3$ are given by
\begin{equation}
\mathrm{selfmix}_{L_{\rm in} \to L_{\rm out}}(\mathbf{x})^{(l_3)} =
\begin{cases}
 \mathbf{k}_{l_3}\circ\mathbf{x}^{(l_3)} + \sum_{l_1=0}^{L_{\rm in}}\sum_{l_2=l_1+1}^{L_{\rm in}} \mathbf{s}_{l_3,l_2,l_1}\circ\left(\mathbf{x}^{(l_1)}\underset{l_3}{\otimes}\mathbf{x}^{(l_2)}\right) & l_3 \leq L_{\rm in}\\
  \sum_{l_1=0}^{L_{\rm in}}\sum_{l_2=l_1+1}^{L_{\rm in}} \mathbf{s}_{l_3,l_2,l_1}\circ\left(\mathbf{x}^{(l_1)}\underset{l_3}{\otimes}\mathbf{x}^{(l_2)}\right) & l_3 > L_{\rm in}
\end{cases}\,.
\label{eq:selfmix_layer}
\end{equation}
Here, $\mathbf{k},\mathbf{s} \in \mathbb{R}^{F}$ are learnable coefficients and the subscripts are used to distinguish independent parameters $\mathbf{k},\mathbf{s}$ for different degrees: In total, a selfmix layer has $L_{\rm out}+1$ different $\mathbf{k}_{l_3}$ (one for each possible value of $l_3\in\{ 0,\dots,L_{\rm out}\}$) and $(L_{\rm out}+1)\frac{L_{\rm in} (L_{\rm in}+1)}{2}$ different $\mathbf{s}_{l_3,l_2,l_1}$ (one for each valid combination of $l_3,l_2,l_1$). 
\paragraph{Spherical linear layers} Spherical linear layers are a combination of linear (Eq.~\ref{eq:linear_layer}) and selfmix (Eq.~\ref{eq:selfmix_layer}) layers given by
\begin{equation}
\mathrm{sphlinear}_{L_{\rm in} \to L_{\rm out},F_{\rm in} \to F_{\rm out}}(\mathbf{x}) =
\mathrm{linear}_{F_{\rm in} \to F_{\rm out}}\left(\mathrm{selfmix_{L_{\rm in} \to L_{\rm out}}(\mathbf{x})}\right)\,.
\label{eq:spherical_linear_layer}
\end{equation}
Chaining both operations allows arbitrary combinations across feature channels and degrees while still preserving rotational equivariance. In principle, whenever $L_{\rm in} = L_{\rm out}$, selfmix layers are not strictly necessary and Eq.~\ref{eq:spherical_linear_layer} may be replaced by Eq.~\ref{eq:linear_layer} for a boost in computational efficiency and reduction of memory footprint. However, this comes at the cost of reduced accuracy (see Section~\ref{Ssec:ablation_studies} in the supplement for details).

\paragraph{Residual blocks} are modules consisting of two sequential spherical linear layers (see Eq.~\ref{eq:spherical_linear_layer}) and activation functions (see Eqs.~\ref{eq:activation_function1}~and~\ref{eq:activation_function2}) inspired by the pre-activation residual block described in \citep{he2016identity}:
\begin{equation}
\mathrm{residual}(\mathbf{x}) = \mathbf{x} +  \mathrm{sphlinear}_2(\boldsymbol{\sigma}_2(\mathrm{sphlinear}_1(\boldsymbol{\sigma}_1(\mathbf{x}))))\,.
\label{eq:residual_block}
\end{equation}
Here, $F_{\rm in}=F_{\rm out}$ and $L_{\rm in}=L_{\rm out}$ for both spherical linear layers.
 
\paragraph{Pairmix layers} are used to combine a pair of features $\mathbf{x}\in \mathbb{R}^{F\times (L_{x}+1)^2}$ and $\mathbf{y}\in \mathbb{R}^{F\times (L_{y}+1)^2}$ with a scalar $r$ (e.g.\ their Euclidean distance) to generate new features of degree $L_{\rm out}$:
\begin{equation}
\mathrm{pairmix}_{L_{x},L_{y}\to L_{\rm out} } (\mathbf{x}, \mathbf{y}, r)^{(l_3)} =
\sum_{l_1=0}^{L_{x}}\sum_{l_2=0}^{L_{y}} \left(\mathbf{W}_{l_3,l_2,l_1}\mathbf{g}(r)\right)\circ\left(\mathbf{x}^{(l_1)}\underset{l_3}{\otimes}\mathbf{y}^{(l_2)}\right)\,.
\label{eq:pairmix_layer}
\end{equation}
Here, $\mathbf{g}(r) \in \mathbb{R}^{K}$ is the vector $[g_0(r)\ g_1(r)\ \dots\ g_{K-1}(r)]^{\top}$ and $g_k(r)$ are radial basis functions. In this work, exponential Bernstein polynomials \citep{unke2021spookynet} are used (see Section~\ref{Ssubsec:bernstein_polynomials} and Eq.~\ref{Seq:bernstein_basis_function} in the supplement). 
The weight matrices $\mathbf{W} \in \mathbb{R}^{F\times K}$ allow to learn radial functions as linear combinations of the basis functions $g_k(r)$ and subscripts are used to distinguish independent weights for different combinations of $l_1,l_2,l_3$ (in total, there are $(L_x+1)(L_y+1)(L_{\rm out}+1)$ possible combinations).  

\paragraph{Interaction blocks} use message-passing to model interactions between the features $\mathbf{c} \in \mathbb{R}^{F\times(L+1)^2}$ of a central point $i$ with features $\mathbf{n} \in \mathbb{R}^{F\times(L+1)^2}$ of neighboring points $j$ within a local environment:
\begin{equation}
\begin{aligned}
\mathbf{a}(\mathbf{x},\mathbf{r})^{(l)} &= \mathbf{x}^0\circ\left(\mathbf{W}_{l}\mathbf{g}(\lVert \mathbf{r}\rVert)\right)\circ\mathrm{sphlinear}_{L\to L,1 \to F}(\mathbf{Y}(\mathbf{r}))^{(l)}\,,\\
\mathbf{b}(\mathbf{x},\mathbf{r})^{(l)} &= \mathrm{pairmix}_{L,L\to L}(\mathbf{x},\mathrm{sphlinear}_{L\to L,1 \to F}\left(\mathbf{Y}(\mathbf{r})\right),\lVert \mathbf{r}\rVert)^{(l)}\,,\\
\mathrm{interaction}(\mathbf{c}, \mathbf{n},  \mathbf{r})^{(l)}_i &= \mathbf{c}_i^{(l)} + \sum_{j\neq i}\left(\mathbf{a}(\mathbf{n}_j,\mathbf{r}_{ij})^{(l)} + \mathbf{b}(\mathbf{n}_j,\mathbf{r}_{ij})^{(l)}\right)\,.
\end{aligned}
\label{eq:interaction_block}
\end{equation}
Here $\mathbf{r}_{ij}$ is the distance vector $\mathbf{r}_{ij} = \mathbf{r}_j - \mathbf{r}_i$ between the positions $\mathbf{r}_i, \mathbf{r}_j$ of $i$~and~$j$. The radial basis function expansion $\mathbf{g}$ is the same as in $\mathrm{pairmix}$ layers and $\mathbf{W} \in \mathbb{R}^{F\times K}$ are independent weight matrices for each degree $l$. Since geometric information enters Eq.~\ref{eq:interaction_block} via relative distance vectors $\mathbf{r}_{ij}$ expanded in a spherical harmonics basis, interactions blocks are equivariant with respect to the SE(3) group of roto-translations.

\subsection{PhiSNet architecture}
\label{subsec:architecture_overview}
PhiSNet takes as inputs nuclear charges $Z$ and positions $\mathbf{r}$ of $N$ atoms, which are used to construct equivariant feature representations encoding information about the chemical environment of each atom. These features are then further transformed and used to predict the entries of the Hamiltonian matrix, see below. An overview over the complete architecture is shown in Fig.~\ref{fig:overview}B and more detailed diagrams of individual building blocks are given in Fig.~\ref{Sfig:nn_architecture} (see Section~\ref{Ssec:ablation_studies} of the supplement for ablation studies that explore the impact of possible simplifications of the PhiSNet architecture on prediction accuracy).

\paragraph{Atomic feature representations}
An embedding layer produces initial atomic feature representations $\mathbf{x}$ from the nuclear charges $Z$ according to
\begin{equation}
\mathrm{embedding}(Z)^{(l)} = 
\begin{cases}
\mathbf{W}\mathbf{d}_Z + \mathbf{b}_Z & l = 0 \\
\mathbf{0} & l > 0
\end{cases}\,,
\label{eq:embedding_layer}
\end{equation}
where $\mathbf{W} \in \mathbb{R}^{F\times 4}$ is a weight matrix and $\mathbf{b}_Z$ element-specific biases with learnable parameters. Here, $\mathbf{d}_Z \in \mathbb{R}^4$ are fixed vectors for each element that contain information about their nuclear charge and ground state electron configuration, similar to the embeddings described in \citep{unke2021spookynet} (see Section~\ref{Ssubsec:embedding} in the supplement for details). The features are then refined by five sequential modules,  each consisting of identical building blocks with independent parameters:
\begin{equation}
\begin{aligned}
\mathbf{t} &= \mathrm{residual}(\mathbf{x})\,, \\
\mathbf{i} &= \mathrm{sphlinear}_{L \to L,F \to F}\left(\boldsymbol{\sigma}\left(\mathrm{residual}( \mathbf{t})\right)\right)\,,  \\
\mathbf{j} &= \mathrm{sphlinear}_{L \to L,F \to F}\left(\boldsymbol{\sigma}\left(\mathrm{residual}( \mathbf{t})\right)\right)\,,  \\
\mathbf{v} &= \mathrm{sphlinear}_{L \to L,F \to F}\left(\boldsymbol{\sigma}\left(\mathrm{residual}(\mathrm{interaction}(\mathbf{i}, \mathbf{j},  \mathbf{r}))\right)\right)\,,  \\
\tilde{\mathbf{x}} &= \mathrm{residual}\left(\mathbf{t} + \mathbf{v}\right)\,, \\
\tilde{\mathbf{y}} &= \mathrm{residual}\left(\tilde{\mathbf{x}}\right)\,.
\end{aligned}
\label{eq:neural_network_module}
\end{equation}

Each module produces two different outputs $\mathbf{\tilde{x}}$ and $\mathbf{\tilde{y}}$. The first output $\mathbf{\tilde{x}}$ serves as input to the next module in the chain (replacing $\mathbf{x}$ in Eq.~\ref{eq:neural_network_module}), whereas the second output $\mathbf{\tilde{y}}$ is summed with the outputs of other modules $m$ to form the final atomic feature representations $\mathbf{f}=\sum_{m}\mathbf{\tilde{y}}_m$ (see Fig.~\ref{Sfig:nn_architecture}A for a visual representation).

\paragraph{Hamiltonian matrix prediction}
The Hamiltonian matrix is constructed block-by-block, with each block corresponding to the interaction between two atoms $i$ and $j$. Diagonal and off-diagonal blocks are treated separately, i.e.\ different atomic pair features are constructed to predict them. Since these transformations also involve interactions with neighboring atoms (similar to interaction blocks), separate representations for central $\mathbf{c}$ and neighboring atoms $\mathbf{n}$ are created from the atomic features $\mathbf{f}$:      
\begin{equation}
\begin{aligned}
\mathbf{c}_i &= \mathrm{residual}(\mathbf{f}_i)\,, \\
\mathbf{n}_j &= \mathrm{residual}(\mathbf{f}_j)\,. \\
\end{aligned}
\label{eq:center_neighbor_features}
\end{equation}

Self-interaction features $\mathbf{f}_{ii}$ (for diagonal blocks) are constructed according to
\begin{equation}
  \mathbf{f}_{ii}^{(l)} = \mathrm{residual}\left(\mathbf{c}_i^{(l)} + \sum_{j\neq i}\left(\mathbf{n}_j^{(l)} \circ \mathbf{W}_l\mathbf{g}(\|\mathbf{r}_{ij}\|)\right)\right)\,,
\label{eq:self_interaction_feats}
\end{equation}
where $\mathbf{r}_{ij}$ is the distance vector between the positions of atoms~$i$~and~$j$, and the radial basis function expansion $\mathbf{g}$ is the same as in $\mathrm{pairmix}$ layers. The matrices $\mathbf{W} \in \mathbb{R}^{F\times K}$ are separate trainable weight matrices for each degree $l$. 

Similarly, pair-interaction features $\mathbf{f}_{ij}$  (for off-diagonal blocks) are obtained by combining the central representations of atoms $i$ and $j$ and interacting them with the neighbors of atom~$i$ according to
\begin{equation}
\mathbf{f}_{ij}^{(l)} = \mathrm{residual}\left(\mathrm{pairmix}(\mathbf{c}_i, \mathbf{c}_j, \|\mathbf{r}_{ij}\|)^{(l)} + \sum_{k\notin \{i,j\}}\left(\mathbf{n}_k^{(l)} \circ \mathbf{W}_l\mathbf{g}(\|\mathbf{r}_{ik}\|)\right)\right)\,.
\label{eq:pair_interaction_feats}
\end{equation}

All blocks $\mathbf{H}_{ii}$ and $\mathbf{H}_{ij}$ of the Hamiltonian matrix, each representing the interaction between two atoms, are themselves composed of smaller blocks corresponding to the interaction between atomic orbitals. Since atomic orbitals are expressed in a spherical harmonics basis, their interactions transform non-trivially (put predictably) under rotations. The correct equivariant behavior of a matrix block $\mathbf{M}^{l_1, l_2}\in \mathbb{R}^{(2l_1 + 1)\times (2l_2 + 1)}$ corresponding to the interaction between orbitals of degree $l_1$ and $l_2$ can be constructed as a sum over matrices obtained from tensor product expansions (Eq.~\ref{eq:tensor_product_expansion}) of irreps $\mathbf{a}$ (collected from specific channels of the pairwise features $\mathbf{f}_{ii}$ or $\mathbf{f}_{ij}$, see below) of all valid degrees $l_3 \in \{|l_2 - l_1|, \dots, \leq l_2 + l_1\}$:
\begin{equation}
\begin{aligned}
  \mathbf{M}^{l_1, l_2} &= \sum_{l_3 = |l_2 - l_1|}^{l_2 + l_1}\overset{l_1,l_2}{\overline{\otimes}}\mathbf{a}^{(l_3)}\,. \\
\end{aligned}
\label{eq:matrix_reconstruction}
\end{equation}

Two sets of indices $I^{\mathrm{self}}$ and $I^{\mathrm{pair}}$ count and keep track of the irreps necessary to construct all required matrices $\mathbf{M}^{l_1, l_2}$. For diagonal blocks $\mathbf{H}_{ii}$, the irreps of a given degree $l$ for the interaction of orbitals $n$ and $m$ of atoms with nuclear charge $Z$ are collected from specific channels of the self-interaction features $\mathbf{f}_{ii}$ via a unique index $I^{\mathrm{self}}(Z, n, m, L)$. Similarly, for off-diagonal blocks $\mathbf{H}_{ij}$, a unique index $I^{\mathrm{pair}}(Z_i, Z_j, n_i, n_j, L)$ selects irreps corresponding to orbitals $n_i$ and $n_j$ of atom pairs with nuclear charges $Z_i$ and $Z_j$ from the pair-interaction features $\mathbf{f}_{ij}$. After all blocks have been constructed, the complete matrix $\tilde{\mathbf{H}}$ is obtained by placing individual blocks at the appropriate positions (based on which atoms and orbitals interact). Finally, a Hamiltonian matrix satisfying the necessary Hermitian symmetry is constructed as $\mathbf{H} = \tilde{\mathbf{H}} + \tilde{\mathbf{H}}^{T}$. This symmetrization guarantees that both pair-features $\mathbf{f}_{ij}$ and $\mathbf{f}_{ji}$ contribute equally to the corresponding off-diagonal blocks and also makes sure that sub-blocks of the Hamiltonian swap positions in the correct way when equivalent atoms are permuted. In cases where multiple Hamiltonian-like matrices need to be predicted, all parameters up to the final residual blocks in Eqs.~\ref{eq:self_interaction_feats}~and~\ref{eq:pair_interaction_feats} are shared. An exception are overlap matrices (see Section~\ref{sec:background}), for which simpler self- and pair-interaction features derived directly from the embeddings are sufficient (see Section~\ref{Ssubsec:overlap_matrix} in the supplement for details). The complete block-wise construction process of the Hamiltonian matrix from irreps is illustrated in Fig.~\ref{Sfig:hamiltonian_composition} for a water molecule with minimal basis set. 

\section{Results and discussion}
\label{sec:results}

To assess the ability of PhiSNet to predict molecular wavefunctions and electronic densities (see Fig.~\ref{fig:overview}C), we train it on Kohn-Sham (the Kohn-Sham matrix takes the role of the Hamiltonian in DFT methods, see Section~\ref{Ssubsec:quantum_chemistry} in the supplement) and overlap matrices for various non-equilibrium configurations of water, ethanol, malondialdehyde, uracil, and aspirin computed at the density functional theory (DFT) level with PBE/def2-SVP. Datasets for all molecules are taken from \citep{schutt2019unifying}, with the exception of aspirin, for which geometries were sampled from the MD17 dataset~\citep{chmiela2018towards} (more details on the datasets, training procedure, and hyperparameter settings can be found in Sections~\ref{Ssec:datasets}~and~\ref{Ssec:training_procedure} of the supplement).

\begin{table}
\label{tab:comparison_with_schnorb}
  \caption{Prediction errors of PhiSNet for various molecules compared to SchNOrb \citep{schutt2019unifying}. In addition to Kohn-Sham $\mathbf{K}$ and overlap $\mathbf{S}$ matrices, we also report errors for energies $\epsilon$ and the cosine similarity between predicted and reference wavefunction $\psi$ for all occupied orbitals. Best results in bold.}
\begin{tabularx}{\textwidth}{c *{5}{>{\centering\arraybackslash}X}}
\toprule
\multirow{2}{*}{Data set} & & $\mathbf{K}$ & $\mathbf{S}$ & $\epsilon$ & $\psi$ \\
 & & [$10^{-6}~\mathrm{E}_{\rm h}$] & [$10^{-6}$] & [$10^{-6}~\mathrm{E}_{\rm h}$] &  \\

\midrule

\multirow{2}{*}{Water} & SchNOrb & 165.4 & 79.1 & 279.3 & \bf 1.00 \\
& PhiSNet & \bf 17.59 & \bf 1.56 & \bf 85.53 & \bf 1.00 \\

\midrule

\multirow{2}{*}{Ethanol} & SchNOrb & 187.4 & 67.8 & 334.4 & \bf 1.00 \\
& PhiSNet & \bf 12.15 & \bf 0.626 & \bf 62.75 & \bf 1.00 \\

\midrule

\multirow{2}{*}{Malondialdehyde} & SchNOrb & 191.1 & 67.3 & 400.6 & 0.99 \\
& PhiSNet & \bf 12.32 & \bf 0.567 & \bf 73.50 & \bf 1.00 \\

\midrule

\multirow{2}{*}{Uracil} & SchNOrb & 227.8 & 82.4 & 1760 & 0.90 \\
& PhiSNet & \bf 10.73 & \bf 0.533 & \bf 84.03 & \bf 1.00 \\

\midrule

\multirow{2}{*}{Aspirin} & SchNOrb & 506.0 & 110 & 48689 & 0.57 \\
& PhiSNet & \bf 12.84 & \bf 0.406 & \bf 176.6 & \bf 0.98 \\

\bottomrule
\end{tabularx}
\end{table}

The results are summarized in Tab.~\ref{tab:comparison_with_schnorb} and compared to the current state-of-the-art given by SchNOrb~\citep{schutt2019unifying}. PhiSNet achieves accuracy improvements up to two orders of magnitude, with the biggest differences arising in larger and more complex molecules like uracil and aspirin. Note that the training process for SchNOrb requires data augmentation via random rotations to approximate the equivariance relation between wavefunction and molecular orientation, while PhiSNet preserves exact equivariance. This not only allows for faster convergence, but also leads to much smaller model sizes, with our model requiring approximately one fifth of the parameters of SchNOrb while providing significantly more accurate results. In addition, PhiSNet provides speedups of over three orders of magnitude compared to DFT calculations (see Section~\ref{Ssubsec:inference_times} for details).

The improved prediction accuracy provided by PhiSNet makes it possible to accurately derive properties such as energies and forces directly from the wavefunction in an end-to-end manner, enabling a number of interesting and novel ML applications for the molecular sciences. As an example, we showcase a transfer-learning application, where a model trained on low accuracy Hartree-Fock (HF) electronic structure calculations is fine-tuned to learn a correction to the wavefunction, such that energies and forces match those obtained via high-level coupled cluster with singles, doubles, and perturbative triple excitations (CCSD(T)) calculations. The CCSD(T) method models the effects of electronic many-body interactions, which are neglected in HF theory, and is often considered to be the ``gold standard'' of quantum chemistry~\citep{unke2020machine}. However, its accuracy comes at a significantly increased computational complexity, going from $\mathcal{O}(N^3)$ for HF to $\mathcal{O}(N^7)$ for CCSD(T) (here, $N$ is the number of basis functions). Thus, a machine-learned correction to HF theory, which mimics the effects of electron correlation in a computationally efficient manner, is a possible way towards novel hybrid methods that rival the accuracy of high level electronic structure calculations and combine the generality and robustness of $\textit{ab initio}$ methods with the efficiency of ML.

After pretraining PhiSNet on HF/cc-pVDZ data for Fock matrices $\mathbf{F}$, core Hamiltonians $\mathbf{H}^{\rm core}$, and overlap matrices $\mathbf{S}$, we fine-tune it on forces computed at the CCSD(T)/cc-pVTZ level, but still keep loss terms for $\mathbf{H}^{\rm core}$ and $\mathbf{S}$ computed with HF/cc-pVDZ (see Section~\ref{Ssubsec:quantum_chemistry} in the supplement for a brief overview of HF theory, where we explain the relevance of $\mathbf{F}$, $\mathbf{H}^{\rm core}$, and $\mathbf{S}$, and how energies and forces are derived from them). This way, the model learns to adapt only the Fock matrix $\mathbf{F}$, which embodies the electron-electron interactions in the HF formalism. Interestingly, very subtle changes to $\mathbf{F}$ seem to be sufficient to approximate the effects of electron correlation, resulting in an ``effective CCSD(T) wavefunction'' that, at first glance, appears to be almost identical to its original HF-level counterpart (see Fig.~\ref{fig:overview}D). Nonetheless, the modified wavefunction reduces the mean absolute errors (MAEs) between energies and forces predicted with PhiSNet to just 79~$\mu$E$_\mathrm{h}$ and 0.85~mE$_\mathrm{h}$a$_0^{-1}$, respectively, compared to the CCSD(T) reference. In contrast, the original HF-level wavefunction leads to MAEs of 4266~$\mu$E$_\mathrm{h}$ and 15~mE$_\mathrm{h}$a$_0^{-1}$ for energies and forces, respectively, i.e.\ prediction errors are reduced by over one order of magnitude with no additional computational overhead. More details on the transfer learning application can be found in Section~\ref{Ssubsec:transfer_learning} of the supplement.

Although PhiSNet predicts electronic properties very accurately, some applications might require \textit{exact} solutions, making existing quantum chemistry methods preferable over predictions of an ML model. Even then, PhiSNet can be used to achieve speedups without any loss of accuracy or robustness: A significant fraction of compute time in HF and DFT calculations must be spent to arrive at a self-consistent solution of Eq.~\ref{eq:schrodinger_matrix_equation}. A good initial guess for the wavefunction can reduce the number of required iterations significantly. We observe a decrease between 56--72\% in the number of iterations, resulting in a total reduction of wall-clock time between 24--40\%, when using PhiSNet-predicted wavefunctions in place of the default guess (see Section~\ref{Ssubsec:initial_guess} in the supplement for more details).

While the results are highly promising, the current approach also has limitations. Due to the fact that all pairwise combinations of atoms have to be considered for constructing the pair-interaction features $\mathbf{f}_{ij}$, the Hamiltonian matrix prediction scales quadratically with the number of atoms. Further, to derive the coefficient matrix $\mathbf{C}$ defining the wavefunction, a generalized eigenvalue problem has to be solved (see Eq.~\ref{eq:schrodinger_matrix_equation}), which scales cubically with the matrix size (the number of basis functions). For these reasons, PhiSNet does not scale well to systems with a very large number of atoms. However, possible extensions of our method could exploit the fact that orbital overlaps between distant atoms are very small, so the entries of corresponding matrix blocks are approximately zero \citep{goedecker2003linear}. Then, only pair-interaction features for non-zero blocks need to be computed and the solution of Eq.~\ref{eq:schrodinger_matrix_equation} can exploit sparsity.

\section{Conclusion}
\label{sec:conclusions}
For learning problems with known invariances, equivariances, symmetries, or other constraints, as is common in physics applications, it is useful to include such properties directly into the model architecture. This effectively reduces the complexity of the learning problem (cf.~\cite{anselmi2016invariance,braun2008relevant}), increasing model performance and decreasing the required amount of training data. While invariance and equivariance properties could also be (approximately) learned, this typically requires much more reference data and/or data augmentation. In contrast, by ``hard coding'' domain knowledge, the learning is constrained to a meaningful submanifold, e.g.\ reflecting rotational equivariance~\citep{marcos2017rotation}, energy conservation, physical laws or symmetries (e.g.~\citep{chmiela2017machine,chmiela2018towards,unke2019physnet,unke2021spookynet,sauceda2021bigdml}), group equivariance (e.g.~\citep{cohen2016group}), graph properties (e.g.~\citep{kondor2018covariant,NEURIPS2019_ea9268cb}) or alike. Thus, known properties do not need to be learned explicitly, because the submanifold where learning takes place already embodies them.  

Our present contribution follows this design principle, specifically, we describe a series of general SE(3)-equivariant operations and building blocks for deep learning architectures operating on geometric point cloud data, which we used here to construct PhiSNet, a novel neural network architecture capable of accurately predicting wavefunctions and electronic densities. Unlike previous models, which need to approximately learn how the wavefunction transforms under molecular rotations and rely on data augmentation, the SE(3)-equivariant\footnote{Note that SE(3)-equivariance is  important (in contrast to just SO(3)-equivariance), because the properties of chemical systems may change substantially when individual atoms are translated relative to each other.} building blocks of our network allow to exactly capture the correct transformation without needing to learn it explicitly. By applying PhiSNet on a range of small to medium-sized molecules, we demonstrated that our model achieves accuracy improvements of up to 
two orders of magnitude compared to the previous state-of-the-art (while at the same time requiring significantly less parameters), and speedups with respect to \textit{ab initio} solutions of over three orders of magnitude. 

For the first time, sufficient accuracy is reached to derive quantum mechanical observations directly from the predicted wavefunction in an end-to-end manner, which also allows to adapt the predicted wavefunction such that it leads to desired physical properties. To showcase such an application, we fine-tuned a model trained on low accuracy wavefunctions to predict properties computed at a much higher level of theory, thereby learning a correction that implicitly captures the effects of many-body electron correlation. This paves the way towards the development of novel semi-empirical methods that are capable of providing highly accurate quantum chemical calculations at a drastically reduced computational cost. In addition, we demonstrated that existing \textit{ab initio} approaches can benefit from guess wavefunctions provided by our method, achieving significant speedups without any loss of accuracy or robustness.

Although we focus on quantum chemistry applications in this contribution, we would like to reiterate that the presented SE(3)-equivariant operations 
are general and can be used to construct other deep learning architectures for geometric point cloud data beyond physics, e.g.~\citep{cohen2016group,marcos2017rotation,kondor2018covariant,NEURIPS2019_b9cfe8b6,NEURIPS2019_ea9268cb}. 

\begin{ack}
 OTU acknowledges funding from the Swiss National Science Foundation (Grant No. P2BSP2\_188147). 
 MB acknowledges support by the Federal Ministry of Education and Research (BMBF) for BIFOLD (01IS18037A).
 KRM was supported in part by the Institute of Information \& Communications Technology Planning \& Evaluation (IITP) grant funded by the Korea Government (No. 2019-0-00079,  Artificial Intelligence Graduate School Program, Korea University), and was partly supported by the German Ministry for Education and Research (BMBF) under Grants 01IS14013A-E, 01GQ1115, 01GQ0850, 01IS18025A and 01IS18037A; the German Research Foundation (DFG) under Grant Math+, EXC 2046/1, Project ID 390685689.
 M. Gastegger works at the BASLEARN Joint Lab for Machine Learning, co-financed by TU Berlin and BASF SE.
 There are no competing interests to declare.
\end{ack}

\bibliography{references}

\begin{thebibliography}{61}
\providecommand{\natexlab}[1]{#1}
\providecommand{\url}[1]{\texttt{#1}}
\expandafter\ifx\csname urlstyle\endcsname\relax
  \providecommand{\doi}[1]{doi: #1}\else
  \providecommand{\doi}{doi: \begingroup \urlstyle{rm}\Url}\fi

\bibitem[Behler and Parrinello(2007)]{behler2007generalized}
J{\"o}rg Behler and Michele Parrinello.
\newblock Generalized neural-network representation of high-dimensional
  potential-energy surfaces.
\newblock \emph{Physical Review Letters}, 98\penalty0 (14):\penalty0 146401,
  2007.

\bibitem[Bart{\'o}k et~al.(2010)Bart{\'o}k, Payne, Kondor, and
  Cs{\'a}nyi]{bartok2010gaussian}
Albert~P. Bart{\'o}k, Mike~C. Payne, Risi Kondor, and G{\'a}bor Cs{\'a}nyi.
\newblock Gaussian approximation potentials: The accuracy of quantum mechanics,
  without the electrons.
\newblock \emph{Physical Review Letters}, 104\penalty0 (13):\penalty0 136403,
  2010.

\bibitem[Smith et~al.(2017)Smith, Isayev, and Roitberg]{smith2017ani}
Justin~S. Smith, Olexandr Isayev, and Adrian~E. Roitberg.
\newblock {ANI-1}: an extensible neural network potential with {DFT} accuracy
  at force field computational cost.
\newblock \emph{Chemical science}, 8\penalty0 (4):\penalty0 3192--3203, 2017.

\bibitem[Chmiela et~al.(2017)Chmiela, Tkatchenko, Sauceda, Poltavsky,
  Sch{\"u}tt, and M{\"u}ller]{chmiela2017machine}
Stefan Chmiela, Alexandre Tkatchenko, Huziel~E. Sauceda, Igor Poltavsky,
  Kristof~T. Sch{\"u}tt, and Klaus-Robert M{\"u}ller.
\newblock Machine learning of accurate energy-conserving molecular force
  fields.
\newblock \emph{Science Advances}, 3\penalty0 (5):\penalty0 e1603015, 2017.

\bibitem[Chmiela et~al.(2018)Chmiela, Sauceda, M{\"u}ller, and
  Tkatchenko]{chmiela2018towards}
Stefan Chmiela, Huziel~E. Sauceda, Klaus-Robert M{\"u}ller, and Alexandre
  Tkatchenko.
\newblock Towards exact molecular dynamics simulations with machine-learned
  force fields.
\newblock \emph{Nature Communications}, 9\penalty0 (1):\penalty0 3887, 2018.

\bibitem[Christensen et~al.(2019)Christensen, Faber, and von
  Lilienfeld]{christensen2019operators}
Anders~S. Christensen, Felix~A. Faber, and O.~Anatole von Lilienfeld.
\newblock Operators in quantum machine learning: Response properties in
  chemical space.
\newblock \emph{Journal of Physical Chemistry}, 150\penalty0 (6):\penalty0
  064105, 2019.

\bibitem[Batzner et~al.(2021)Batzner, Smidt, Sun, Mailoa, Kornbluth, Molinari,
  and Kozinsky]{Batzner2021_e3nn_molecular_dynamics}
Simon Batzner, Tess~E. Smidt, Lixin Sun, Jonathan~P. Mailoa, Mordechai
  Kornbluth, Nicola Molinari, and Boris Kozinsky.
\newblock {SE(3)}-equivariant graph neural networks for data-efficient and
  accurate interatomic potentials.
\newblock \emph{arXiv preprint arXiv:2101.03164}, 2021.

\bibitem[Unke et~al.(2021{\natexlab{a}})Unke, Chmiela, Sauceda, Gastegger,
  Poltavsky, Sch{\"u}tt, Tkatchenko, and M{\"u}ller]{unke2020machine}
Oliver~T. Unke, Stefan Chmiela, Huziel~E. Sauceda, Michael Gastegger, Igor
  Poltavsky, Kristof~T. Sch{\"u}tt, Alexandre Tkatchenko, and Klaus-Robert
  M{\"u}ller.
\newblock Machine learning force fields.
\newblock \emph{Chemical Reviews}, 121\penalty0 (16):\penalty0 10142–10186,
  2021{\natexlab{a}}.

\bibitem[Snyder et~al.(2012)Snyder, Rupp, Hansen, M{\"u}ller, and
  Burke]{snyder2012finding}
John~C. Snyder, Matthias Rupp, Katja Hansen, Klaus-Robert M{\"u}ller, and
  Kieron Burke.
\newblock Finding density functionals with machine learning.
\newblock \emph{Physical Review Letters}, 108\penalty0 (25):\penalty0 253002,
  2012.

\bibitem[Brockherde et~al.(2017)Brockherde, Vogt, Li, Tuckerman, Burke, and
  M{\"u}ller]{brockherde2017bypassing}
Felix Brockherde, Leslie Vogt, Li~Li, Mark~E. Tuckerman, Kieron Burke, and
  Klaus-Robert M{\"u}ller.
\newblock Bypassing the {Kohn-Sham} equations with machine learning.
\newblock \emph{Nature Communications}, 8:\penalty0 872, 2017.

\bibitem[Ryczko et~al.(2018)Ryczko, Strubbe, and Tamblyn]{ryczko2018deep}
Kevin Ryczko, David Strubbe, and Isaac Tamblyn.
\newblock Deep learning and density functional theory.
\newblock \emph{arXiv preprint arXiv:1811.08928}, 2018.

\bibitem[Grisafi et~al.(2018{\natexlab{a}})Grisafi, Wilkins, Meyer, Fabrizio,
  Corminboeuf, and Ceriotti]{grisafi2018transferable}
Andrea Grisafi, David~M. Wilkins, Benjamin A.~R. Meyer, Alberto Fabrizio,
  Clemence Corminboeuf, and Michele Ceriotti.
\newblock A transferable machine-learning model of the electron density.
\newblock \emph{arXiv preprint arXiv:1809.05349}, 2018{\natexlab{a}}.

\bibitem[Nagai et~al.(2020)Nagai, Akashi, and Sugino]{nagai2020completing}
Ryo Nagai, Ryosuke Akashi, and Osamu Sugino.
\newblock Completing density functional theory by machine learning hidden
  messages from molecules.
\newblock \emph{npj Computational Materials}, 6\penalty0 (1):\penalty0 1--8,
  2020.

\bibitem[Bogojeski et~al.(2020)Bogojeski, Vogt-Maranto, Tuckerman, M{\"u}ller,
  and Burke]{bogojeski2020quantum}
Mihail Bogojeski, Leslie Vogt-Maranto, Mark~E. Tuckerman, Klaus-Robert
  M{\"u}ller, and Kieron Burke.
\newblock Quantum chemical accuracy from density functional approximations via
  machine learning.
\newblock \emph{Nature communications}, 11:\penalty0 5223, 2020.

\bibitem[Rupp et~al.(2012)Rupp, Tkatchenko, M{\"u}ller, and von
  Lilienfeld]{rupp2012fast}
Matthias Rupp, Alexandre Tkatchenko, Klaus-Robert M{\"u}ller, and O.~Anatole
  von Lilienfeld.
\newblock Fast and accurate modeling of molecular atomization energies with
  machine learning.
\newblock \emph{Physical Review Letters}, 108\penalty0 (5):\penalty0 058301,
  2012.

\bibitem[Montavon et~al.(2013)Montavon, Rupp, Gobre, Vazquez-Mayagoitia,
  Hansen, Tkatchenko, M{\"{u}}ller, and von Lilienfeld]{Montavon2013a}
Gr{\'{e}}goire Montavon, Matthias Rupp, Vivekanand Gobre, Alvaro
  Vazquez-Mayagoitia, Katja Hansen, Alexandre Tkatchenko, Klaus-Robert
  M{\"{u}}ller, and O.~Anatole von Lilienfeld.
\newblock {Machine learning of molecular electronic properties in chemical
  compound space}.
\newblock \emph{New Journal of Physics}, 15\penalty0 (9):\penalty0 095003,
  2013.

\bibitem[De et~al.(2016)De, Bart{\'o}k, Cs{\'a}nyi, and
  Ceriotti]{de2016comparing}
Sandip De, Albert~P Bart{\'o}k, G{\'a}bor Cs{\'a}nyi, and Michele Ceriotti.
\newblock Comparing molecules and solids across structural and alchemical
  space.
\newblock \emph{Physical Chemistry Chemical Physics}, 18\penalty0
  (20):\penalty0 13754--13769, 2016.

\bibitem[Sch{\"u}tt et~al.(2017{\natexlab{a}})Sch{\"u}tt, Arbabzadah, Chmiela,
  M{\"u}ller, and Tkatchenko]{schutt2017quantum}
Kristof~T. Sch{\"u}tt, Farhad Arbabzadah, Stefan Chmiela, Klaus-Robert
  M{\"u}ller, and Alexandre Tkatchenko.
\newblock Quantum-chemical insights from deep tensor neural networks.
\newblock \emph{Nature Communications}, 8:\penalty0 13890, 2017{\natexlab{a}}.

\bibitem[Sch{\"u}tt et~al.(2017{\natexlab{b}})Sch{\"u}tt, Kindermans, Sauceda,
  Chmiela, Tkatchenko, and M{\"u}ller]{schutt2017schnet}
Kristof~T. Sch{\"u}tt, Pieter-Jan Kindermans, Huziel~E. Sauceda, Stefan
  Chmiela, Alexandre Tkatchenko, and Klaus-Robert M{\"u}ller.
\newblock Sch{N}et: A continuous-filter convolutional neural network for
  modeling quantum interactions.
\newblock In \emph{Advances in Neural Information Processing Systems}, pages
  991--1001, 2017{\natexlab{b}}.

\bibitem[Sch{\"u}tt et~al.(2018)Sch{\"u}tt, Sauceda, Kindermans, Tkatchenko,
  and M{\"u}ller]{schutt2018schnet}
Kristof~T. Sch{\"u}tt, Huziel~E. Sauceda, Pieter-Jan Kindermans, Alexandre
  Tkatchenko, and Klaus-Robert M{\"u}ller.
\newblock {SchNet} -- a deep learning architecture for molecules and materials.
\newblock \emph{The Journal of Chemical Physics}, 148\penalty0 (24):\penalty0
  241722, 2018.

\bibitem[Pronobis et~al.(2018)Pronobis, Tkatchenko, and
  M{\"u}ller]{pronobis2018many}
Wiktor Pronobis, Alexandre Tkatchenko, and Klaus-Robert M{\"u}ller.
\newblock Many-body descriptors for predicting molecular properties with
  machine learning: Analysis of pairwise and three-body interactions in
  molecules.
\newblock \emph{Journal of Chemical Theory and Computation}, 14\penalty0
  (6):\penalty0 2991--3003, 2018.

\bibitem[Eickenberg et~al.(2018)Eickenberg, Exarchakis, Hirn, Mallat, and
  Thiry]{eickenberg2018solid}
Michael Eickenberg, Georgios Exarchakis, Matthew Hirn, St{\'e}phane Mallat, and
  Louis Thiry.
\newblock Solid harmonic wavelet scattering for predictions of molecule
  properties.
\newblock \emph{The Journal of Chemical Physics}, 148\penalty0 (24):\penalty0
  241732, 2018.

\bibitem[Unke and Meuwly(2019)]{unke2019physnet}
Oliver~T. Unke and Markus Meuwly.
\newblock {PhysNet}: A neural network for predicting energies, forces, dipole
  moments, and partial charges.
\newblock \emph{Journal of Chemical Theory and Computation}, 15\penalty0
  (6):\penalty0 3678--3693, 2019.

\bibitem[von Lilienfeld et~al.(2020)von Lilienfeld, M{\"u}ller, and
  Tkatchenko]{von2020exploring}
O.~Anatole von Lilienfeld, Klaus-Robert M{\"u}ller, and Alexandre Tkatchenko.
\newblock Exploring chemical compound space with quantum-based machine
  learning.
\newblock \emph{Nature Reviews Chemistry}, 4:\penalty0 347--358, 2020.

\bibitem[Chen et~al.(2021)Chen, Andrejevic, Smidt, Ding, Xu, Chi, Nguyen,
  Alatas, Kong, and Li]{Chen2021_e3nn_phonon}
Zhantao Chen, Nina Andrejevic, Tess~E. Smidt, Zhiwei Ding, Qian Xu, Yen-Ting
  Chi, Quynh~T Nguyen, Ahmet Alatas, Jing Kong, and Mingda Li.
\newblock Direct prediction of phonon density of states with {Euclidean} neural
  networks.
\newblock \emph{Advanced Science}, page 2004214, 2021.

\bibitem[Sch{\"u}tt et~al.(2021)Sch{\"u}tt, Unke, and
  Gastegger]{schutt2021equivariant}
Kristof~T. Sch{\"u}tt, Oliver~T. Unke, and Michael Gastegger.
\newblock Equivariant message passing for the prediction of tensorial
  properties and molecular spectra.
\newblock \emph{Proceedings of the 38th International Conference on Machine
  Learning, PMLR 139:9377-9388}, 2021.

\bibitem[Unke et~al.(2021{\natexlab{b}})Unke, Chmiela, Gastegger, Sch{\"u}tt,
  Sauceda, and M{\"u}ller]{unke2021spookynet}
Oliver~T. Unke, Stefan Chmiela, Michael Gastegger, Kristof~T. Sch{\"u}tt,
  Huziel~E. Sauceda, and Klaus-Robert M{\"u}ller.
\newblock {SpookyNet}: Learning force fields with electronic degrees of freedom
  and nonlocal effects.
\newblock \emph{arXiv preprint arXiv:2105.00304}, 2021{\natexlab{b}}.

\bibitem[Keith et~al.(2021)Keith, Vassilev-Galindo, Cheng, Chmiela, Gastegger,
  M{\"u}ller, and Tkatchenko]{keith2021combining}
John~A. Keith, Valentin Vassilev-Galindo, Bingqing Cheng, Stefan Chmiela,
  Michael Gastegger, Klaus-Robert M{\"u}ller, and Alexandre Tkatchenko.
\newblock Combining machine learning and computational chemistry for predictive
  insights into chemical systems.
\newblock \emph{Chemical Reviews}, 121\penalty0 (16):\penalty0 9816–9872,
  2021.
\newblock URL \url{https://pubs.acs.org/doi/abs/10.1021/acs.chemrev.1c00107}.

\bibitem[Gastegger et~al.(2021)Gastegger, Sch{\"u}tt, and
  M{\"u}ller]{gastegger2021machine}
Michael Gastegger, Kristof~T. Sch{\"u}tt, and Klaus-Robert M{\"u}ller.
\newblock Machine learning of solvent effects on molecular spectra and
  reactions.
\newblock \emph{Chemical Science}, 12\penalty0 (34):\penalty0 11473--11483,
  2021.

\bibitem[Qiao et~al.(2020)Qiao, Welborn, Anandkumar, Manby, and
  Miller~III]{qiao2020orbnet}
Zhuoran Qiao, Matthew Welborn, Animashree Anandkumar, Frederick~R. Manby, and
  Thomas~F. Miller~III.
\newblock {OrbNet}: Deep learning for quantum chemistry using symmetry-adapted
  atomic-orbital features.
\newblock \emph{The Journal of Chemical Physics}, 153\penalty0 (12):\penalty0
  124111, 2020.

\bibitem[Carleo and Troyer(2017)]{carleo2017solving}
Giuseppe Carleo and Matthias Troyer.
\newblock Solving the quantum many-body problem with artificial neural
  networks.
\newblock \emph{Science}, 355\penalty0 (6325):\penalty0 602--606, 2017.

\bibitem[Hegde and Bowen(2017)]{hegde2017machine}
Ganesh Hegde and R~Chris Bowen.
\newblock Machine-learned approximations to density functional theory
  {Hamiltonians}.
\newblock \emph{Scientific Reports}, 7:\penalty0 42669, 2017.

\bibitem[Sch{\"u}tt et~al.(2019)Sch{\"u}tt, Gastegger, Tkatchenko, M{\"u}ller,
  and Maurer]{schutt2019unifying}
Kristof~T. Sch{\"u}tt, Michael Gastegger, Alexandre Tkatchenko, Klaus-Robert
  M{\"u}ller, and Reinhard~J. Maurer.
\newblock Unifying machine learning and quantum chemistry with a deep neural
  network for molecular wavefunctions.
\newblock \emph{Nature Communications}, 10:\penalty0 5024, 2019.

\bibitem[Li et~al.(2021)Li, Wang, Zou, Ye, Duan, and Xu]{li2021deep}
He~Li, Zun Wang, Nianlong Zou, Meng Ye, Wenhui Duan, and Yong Xu.
\newblock Deep neural network representation of density functional theory
  {Hamiltonian}.
\newblock \emph{arXiv preprint arXiv:2104.03786}, 2021.

\bibitem[Montavon et~al.(2012)Montavon, Hansen, Fazli, Rupp, Biegler, Ziehe,
  Tkatchenko, Lilienfeld, and M{\"u}ller]{montavon2012learning}
Gr{\'e}goire Montavon, Katja Hansen, Siamac Fazli, Matthias Rupp, Franziska
  Biegler, Andreas Ziehe, Alexandre Tkatchenko, Anatole Lilienfeld, and
  Klaus-Robert M{\"u}ller.
\newblock Learning invariant representations of molecules for atomization
  energy prediction.
\newblock \emph{Advances in neural information processing systems},
  25:\penalty0 440--448, 2012.

\bibitem[Gastegger et~al.(2020)Gastegger, McSloy, Luya, Sch{\"u}tt, and
  Maurer]{gastegger2020deep}
Michael Gastegger, Adam McSloy, M~Luya, Kristof~T. Sch{\"u}tt, and Reinhard~J.
  Maurer.
\newblock A deep neural network for molecular wave functions in quasi-atomic
  minimal basis representation.
\newblock \emph{The Journal of Chemical Physics}, 153\penalty0 (4):\penalty0
  044123, 2020.

\bibitem[Cohen and Welling(2016)]{cohen2016group}
Taco~S. Cohen and Max Welling.
\newblock Group equivariant convolutional networks.
\newblock In \emph{International conference on machine learning}, pages
  2990--2999. PMLR, 2016.

\bibitem[Marcos et~al.(2017)Marcos, Volpi, Komodakis, and
  Tuia]{marcos2017rotation}
Diego Marcos, Michele Volpi, Nikos Komodakis, and Devis Tuia.
\newblock Rotation equivariant vector field networks.
\newblock In \emph{Proceedings of the IEEE International Conference on Computer
  Vision}, pages 5048--5057, 2017.

\bibitem[Cohen et~al.(2018)Cohen, Geiger, K{\"o}hler, and
  Welling]{cohen2018spherical}
Taco~S Cohen, Mario Geiger, Jonas K{\"o}hler, and Max Welling.
\newblock Spherical {CNNs}.
\newblock \emph{arXiv preprint arXiv:1801.10130}, 2018.

\bibitem[Kondor et~al.(2018{\natexlab{a}})Kondor, Son, Pan, Anderson, and
  Trivedi]{kondor2018covariant}
Risi Kondor, Hy~Truong Son, Horace Pan, Brandon Anderson, and Shubhendu
  Trivedi.
\newblock Covariant compositional networks for learning graphs.
\newblock \emph{arXiv preprint arXiv:1801.02144}, 2018{\natexlab{a}}.

\bibitem[Weiler et~al.(2018)Weiler, Geiger, Welling, Boomsma, and
  Cohen]{weiler20183d}
Maurice Weiler, Mario Geiger, Max Welling, Wouter Boomsma, and Taco Cohen.
\newblock {3D} steerable {CNNs}: Learning rotationally equivariant features in
  volumetric data.
\newblock \emph{arXiv preprint arXiv:1807.02547}, 2018.

\bibitem[Thomas et~al.(2018)Thomas, Smidt, Kearnes, Yang, Li, Kohlhoff, and
  Riley]{thomas2018tensor}
Nathaniel Thomas, Tess~E. Smidt, Steven Kearnes, Lusann Yang, Li~Li, Kai
  Kohlhoff, and Patrick Riley.
\newblock Tensor field networks: Rotation-and translation-equivariant neural
  networks for {3D} point clouds.
\newblock \emph{arXiv preprint arXiv:1802.08219}, 2018.

\bibitem[Grisafi et~al.(2018{\natexlab{b}})Grisafi, Wilkins, Cs\'anyi, and
  Ceriotti]{Grisafi2018}
Andrea Grisafi, David~M. Wilkins, G\'abor Cs\'anyi, and Michele Ceriotti.
\newblock Symmetry-adapted machine learning for tensorial properties of
  atomistic systems.
\newblock \emph{Physical Review Letters}, 120:\penalty0 036002, Jan
  2018{\natexlab{b}}.
\newblock \doi{10.1103/PhysRevLett.120.036002}.

\bibitem[Hy et~al.(2018)Hy, Trivedi, Pan, Anderson, and
  Kondor]{hy2018predicting}
Truong~Son Hy, Shubhendu Trivedi, Horace Pan, Brandon~M. Anderson, and Risi
  Kondor.
\newblock Predicting molecular properties with covariant compositional
  networks.
\newblock \emph{The Journal of Chemical Physics}, 148\penalty0 (24):\penalty0
  241745, 2018.

\bibitem[Cohen et~al.(2019{\natexlab{a}})Cohen, Geiger, and
  Weiler]{NEURIPS2019_b9cfe8b6}
Taco~S. Cohen, Mario Geiger, and Maurice Weiler.
\newblock A general theory of equivariant {CNNs} on homogeneous spaces.
\newblock In \emph{Advances in Neural Information Processing Systems},
  volume~32. Curran Associates, Inc., 2019{\natexlab{a}}.

\bibitem[Keriven and Peyr\'{e}(2019)]{NEURIPS2019_ea9268cb}
Nicolas Keriven and Gabriel Peyr\'{e}.
\newblock Universal invariant and equivariant graph neural networks.
\newblock In \emph{Advances in Neural Information Processing Systems},
  volume~32. Curran Associates, Inc., 2019.

\bibitem[Cohen et~al.(2019{\natexlab{b}})Cohen, Weiler, Kicanaoglu, and
  Welling]{cohen2019gauge}
Taco Cohen, Maurice Weiler, Berkay Kicanaoglu, and Max Welling.
\newblock Gauge equivariant convolutional networks and the icosahedral {CNN}.
\newblock In \emph{International Conference on Machine Learning}, pages
  1321--1330. PMLR, 2019{\natexlab{b}}.

\bibitem[Fuchs et~al.(2020)Fuchs, Worrall, Fischer, and Welling]{fuchs2020se}
Fabian~B. Fuchs, Daniel~E. Worrall, Volker Fischer, and Max Welling.
\newblock {SE(3)}-transformers: {3D} roto-translation equivariant attention
  networks.
\newblock \emph{arXiv preprint arXiv:2006.10503}, 2020.

\bibitem[Nigam et~al.(2021)Nigam, Willatt, and Ceriotti]{nigam2021equivariant}
Jigyasa Nigam, Michael Willatt, and Michele Ceriotti.
\newblock Equivariant representations for molecular {Hamiltonians} and
  {N-center} atomic-scale properties.
\newblock \emph{arXiv preprint arXiv:2109.12083}, 2021.

\bibitem[Schechter(2012)]{schechter2012operator}
Martin Schechter.
\newblock \emph{Operator methods in quantum mechanics}.
\newblock Elsevier, 2012.

\bibitem[Gilmer et~al.(2017)Gilmer, Schoenholz, Riley, Vinyals, and
  Dahl]{gilmer2017neural}
Justin Gilmer, Samuel~S. Schoenholz, Patrick~F. Riley, Oriol Vinyals, and
  George~E. Dahl.
\newblock Neural message passing for quantum chemistry.
\newblock In \emph{International Conference on Machine Learning}, pages
  1263--1272. PMLR, 2017.

\bibitem[Hendrycks and Gimpel(2016)]{hendrycks2016gaussian}
Dan Hendrycks and Kevin Gimpel.
\newblock Gaussian error linear units {(GELUs)}.
\newblock \emph{arXiv preprint arXiv:1606.08415}, 2016.

\bibitem[Elfwing et~al.(2018)Elfwing, Uchibe, and Doya]{elfwing2018sigmoid}
Stefan Elfwing, Eiji Uchibe, and Kenji Doya.
\newblock Sigmoid-weighted linear units for neural network function
  approximation in reinforcement learning.
\newblock \emph{Neural Netw.}, 107:\penalty0 3--11, 2018.

\bibitem[Ramachandran et~al.(2017)Ramachandran, Zoph, and
  Le]{ramachandran2017searching}
Prajit Ramachandran, Barret Zoph, and Quoc~V Le.
\newblock Searching for activation functions.
\newblock \emph{arXiv preprint arXiv:1710.05941}, 2017.

\bibitem[Varshalovich et~al.(1988)Varshalovich, Moskalev, and
  Khersonskii]{varshalovich1988quantum}
D.~A. Varshalovich, A.~N. Moskalev, and V.~K. Khersonskii.
\newblock \emph{Quantum theory of angular momentum}.
\newblock World Scientific, 1988.

\bibitem[Kondor et~al.(2018{\natexlab{b}})Kondor, Lin, and
  Trivedi]{kondor2018clebsch}
Risi Kondor, Zhen Lin, and Shubhendu Trivedi.
\newblock Clebsch-gordan nets: a fully fourier space spherical convolutional
  neural network.
\newblock \emph{arXiv preprint arXiv:1806.09231}, 2018{\natexlab{b}}.

\bibitem[He et~al.(2016)He, Zhang, Ren, and Sun]{he2016identity}
Kaiming He, Xiangyu Zhang, Shaoqing Ren, and Jian Sun.
\newblock Identity mappings in deep residual networks.
\newblock In \emph{European conference on computer vision}, pages 630--645.
  Springer, 2016.

\bibitem[Goedecker and Scuserza(2003)]{goedecker2003linear}
Stefan Goedecker and GE~Scuserza.
\newblock Linear scaling electronic structure methods in chemistry and physics.
\newblock \emph{Computing in Science \& Engineering}, 5\penalty0 (4):\penalty0
  14--21, 2003.

\bibitem[Anselmi et~al.(2016)Anselmi, Rosasco, and
  Poggio]{anselmi2016invariance}
Fabio Anselmi, Lorenzo Rosasco, and Tomaso Poggio.
\newblock {On invariance and selectivity in representation learning}.
\newblock \emph{Information and Inference: A Journal of the IMA}, 5\penalty0
  (2):\penalty0 134--158, 2016.
\newblock \doi{10.1093/imaiai/iaw009}.

\bibitem[Braun et~al.(2008)Braun, Buhmann, and M{\"u}ller]{braun2008relevant}
Mikio~L. Braun, Joachim~M. Buhmann, and Klaus-Robert M{\"u}ller.
\newblock On relevant dimensions in kernel feature spaces.
\newblock \emph{The Journal of Machine Learning Research}, 9(62):\penalty0
  1875--1908, 2008.

\bibitem[Sauceda et~al.(2021)Sauceda, G{\'a}lvez-Gonz{\'a}lez, Chmiela,
  Paz-Borb{\'o}n, M{\"u}ller, and Tkatchenko]{sauceda2021bigdml}
Huziel~E. Sauceda, Luis~E. G{\'a}lvez-Gonz{\'a}lez, Stefan Chmiela,
  Lauro~Oliver Paz-Borb{\'o}n, Klaus-Robert M{\"u}ller, and Alexandre
  Tkatchenko.
\newblock {BIGDML}: Towards exact machine learning force fields for materials.
\newblock \emph{arXiv preprint arXiv:2106.04229}, 2021.

\end{thebibliography}


\begin{thebibliography}{34}
\providecommand{\natexlab}[1]{#1}
\providecommand{\url}[1]{\texttt{#1}}
\expandafter\ifx\csname urlstyle\endcsname\relax
  \providecommand{\doi}[1]{doi: #1}\else
  \providecommand{\doi}{doi: \begingroup \urlstyle{rm}\Url}\fi

\bibitem[Cohen and Welling(2016{\natexlab{a}})]{cohen2016group}
Taco~S. Cohen and Max Welling.
\newblock Group equivariant convolutional networks.
\newblock In \emph{International conference on machine learning}, pages
  2990--2999. PMLR, 2016{\natexlab{a}}.

\bibitem[Marcos et~al.(2017)Marcos, Volpi, Komodakis, and
  Tuia]{marcos2017rotation}
Diego Marcos, Michele Volpi, Nikos Komodakis, and Devis Tuia.
\newblock Rotation equivariant vector field networks.
\newblock In \emph{Proceedings of the IEEE International Conference on Computer
  Vision}, pages 5048--5057, 2017.

\bibitem[Cohen and Welling(2016{\natexlab{b}})]{cohen2016steerable}
Taco~S. Cohen and Max Welling.
\newblock Steerable cnns.
\newblock \emph{arXiv preprint arXiv:1612.08498}, 2016{\natexlab{b}}.

\bibitem[Worrall et~al.(2017)Worrall, Garbin, Turmukhambetov, and
  Brostow]{worrall2017harmonic}
Daniel~E Worrall, Stephan~J Garbin, Daniyar Turmukhambetov, and Gabriel~J
  Brostow.
\newblock Harmonic networks: Deep translation and rotation equivariance.
\newblock In \emph{Proceedings of the IEEE Conference on Computer Vision and
  Pattern Recognition}, pages 5028--5037, 2017.

\bibitem[Grisafi et~al.(2018)Grisafi, Wilkins, Cs\'anyi, and
  Ceriotti]{Grisafi2018}
Andrea Grisafi, David~M. Wilkins, G\'abor Cs\'anyi, and Michele Ceriotti.
\newblock Symmetry-adapted machine learning for tensorial properties of
  atomistic systems.
\newblock \emph{Physical Review Letters}, 120:\penalty0 036002, Jan 2018.
\newblock \doi{10.1103/PhysRevLett.120.036002}.

\bibitem[Cohen et~al.(2018)Cohen, Geiger, K{\"o}hler, and
  Welling]{cohen2018spherical}
Taco~S Cohen, Mario Geiger, Jonas K{\"o}hler, and Max Welling.
\newblock Spherical {CNNs}.
\newblock \emph{arXiv preprint arXiv:1801.10130}, 2018.

\bibitem[Thomas et~al.(2018)Thomas, Smidt, Kearnes, Yang, Li, Kohlhoff, and
  Riley]{thomas2018tensor}
Nathaniel Thomas, Tess~E. Smidt, Steven Kearnes, Lusann Yang, Li~Li, Kai
  Kohlhoff, and Patrick Riley.
\newblock Tensor field networks: Rotation-and translation-equivariant neural
  networks for {3D} point clouds.
\newblock \emph{arXiv preprint arXiv:1802.08219}, 2018.

\bibitem[Weiler et~al.(2018)Weiler, Geiger, Welling, Boomsma, and
  Cohen]{weiler20183d}
Maurice Weiler, Mario Geiger, Max Welling, Wouter Boomsma, and Taco Cohen.
\newblock {3D} steerable {CNNs}: Learning rotationally equivariant features in
  volumetric data.
\newblock \emph{arXiv preprint arXiv:1807.02547}, 2018.

\bibitem[Kondor et~al.(2018{\natexlab{a}})Kondor, Lin, and
  Trivedi]{kondor2018clebsch}
Risi Kondor, Zhen Lin, and Shubhendu Trivedi.
\newblock Clebsch-gordan nets: a fully fourier space spherical convolutional
  neural network.
\newblock \emph{arXiv preprint arXiv:1806.09231}, 2018{\natexlab{a}}.

\bibitem[Kondor et~al.(2018{\natexlab{b}})Kondor, Son, Pan, Anderson, and
  Trivedi]{kondor2018covariant}
Risi Kondor, Hy~Truong Son, Horace Pan, Brandon Anderson, and Shubhendu
  Trivedi.
\newblock Covariant compositional networks for learning graphs.
\newblock \emph{arXiv preprint arXiv:1801.02144}, 2018{\natexlab{b}}.

\bibitem[Hy et~al.(2018)Hy, Trivedi, Pan, Anderson, and
  Kondor]{hy2018predicting}
Truong~Son Hy, Shubhendu Trivedi, Horace Pan, Brandon~M. Anderson, and Risi
  Kondor.
\newblock Predicting molecular properties with covariant compositional
  networks.
\newblock \emph{The Journal of Chemical Physics}, 148\penalty0 (24):\penalty0
  241745, 2018.

\bibitem[Cohen et~al.(2019{\natexlab{a}})Cohen, Weiler, Kicanaoglu, and
  Welling]{cohen2019gauge}
Taco Cohen, Maurice Weiler, Berkay Kicanaoglu, and Max Welling.
\newblock Gauge equivariant convolutional networks and the icosahedral {CNN}.
\newblock In \emph{International Conference on Machine Learning}, pages
  1321--1330. PMLR, 2019{\natexlab{a}}.

\bibitem[Keriven and Peyr\'{e}(2019)]{NEURIPS2019_ea9268cb}
Nicolas Keriven and Gabriel Peyr\'{e}.
\newblock Universal invariant and equivariant graph neural networks.
\newblock In \emph{Advances in Neural Information Processing Systems},
  volume~32. Curran Associates, Inc., 2019.

\bibitem[Cohen et~al.(2019{\natexlab{b}})Cohen, Geiger, and
  Weiler]{NEURIPS2019_b9cfe8b6}
Taco~S. Cohen, Mario Geiger, and Maurice Weiler.
\newblock A general theory of equivariant {CNNs} on homogeneous spaces.
\newblock In \emph{Advances in Neural Information Processing Systems},
  volume~32. Curran Associates, Inc., 2019{\natexlab{b}}.

\bibitem[Finzi et~al.(2020)Finzi, Stanton, Izmailov, and
  Wilson]{finzi2020generalizing}
Marc Finzi, Samuel Stanton, Pavel Izmailov, and Andrew~Gordon Wilson.
\newblock Generalizing convolutional neural networks for equivariance to lie
  groups on arbitrary continuous data.
\newblock In \emph{International Conference on Machine Learning}, pages
  3165--3176. PMLR, 2020.

\bibitem[Fuchs et~al.(2020)Fuchs, Worrall, Fischer, and Welling]{fuchs2020se}
Fabian~B. Fuchs, Daniel~E. Worrall, Volker Fischer, and Max Welling.
\newblock {SE(3)}-transformers: {3D} roto-translation equivariant attention
  networks.
\newblock \emph{arXiv preprint arXiv:2006.10503}, 2020.

\bibitem[Vaswani et~al.(2017)Vaswani, Shazeer, Parmar, Uszkoreit, Jones, Gomez,
  Kaiser, and Polosukhin]{vaswani2017attention}
Ashish Vaswani, Noam Shazeer, Niki Parmar, Jakob Uszkoreit, Llion Jones,
  Aidan~N Gomez, {\L}ukasz Kaiser, and Illia Polosukhin.
\newblock Attention is all you need.
\newblock In \emph{Advances in neural information processing systems}, pages
  5998--6008, 2017.

\bibitem[Gilmer et~al.(2017)Gilmer, Schoenholz, Riley, Vinyals, and
  Dahl]{gilmer2017neural}
Justin Gilmer, Samuel~S. Schoenholz, Patrick~F. Riley, Oriol Vinyals, and
  George~E. Dahl.
\newblock Neural message passing for quantum chemistry.
\newblock In \emph{International Conference on Machine Learning}, pages
  1263--1272. PMLR, 2017.

\bibitem[Anderson et~al.(2019)Anderson, Hy, and Kondor]{anderson2019cormorant}
Brandon Anderson, Truong-Son Hy, and Risi Kondor.
\newblock Cormorant: Covariant molecular neural networks.
\newblock \emph{arXiv preprint arXiv:1906.04015}, 2019.

\bibitem[Batzner et~al.(2021)Batzner, Smidt, Sun, Mailoa, Kornbluth, Molinari,
  and Kozinsky]{Batzner2021_e3nn_molecular_dynamics}
Simon Batzner, Tess~E. Smidt, Lixin Sun, Jonathan~P. Mailoa, Mordechai
  Kornbluth, Nicola Molinari, and Boris Kozinsky.
\newblock {SE(3)}-equivariant graph neural networks for data-efficient and
  accurate interatomic potentials.
\newblock \emph{arXiv preprint arXiv:2101.03164}, 2021.

\bibitem[Sch{\"u}tt et~al.(2017)Sch{\"u}tt, Kindermans, Sauceda, Chmiela,
  Tkatchenko, and M{\"u}ller]{schutt2017schnet}
Kristof~T. Sch{\"u}tt, Pieter-Jan Kindermans, Huziel~E. Sauceda, Stefan
  Chmiela, Alexandre Tkatchenko, and Klaus-Robert M{\"u}ller.
\newblock Sch{N}et: A continuous-filter convolutional neural network for
  modeling quantum interactions.
\newblock In \emph{Advances in Neural Information Processing Systems}, pages
  991--1001, 2017.

\bibitem[Unke and Meuwly(2019)]{unke2019physnet}
Oliver~T. Unke and Markus Meuwly.
\newblock {PhysNet}: A neural network for predicting energies, forces, dipole
  moments, and partial charges.
\newblock \emph{Journal of Chemical Theory and Computation}, 15\penalty0
  (6):\penalty0 3678--3693, 2019.

\bibitem[Nair and Hinton(2010)]{nair2010rectified}
Vinod Nair and Geoffrey~E Hinton.
\newblock Rectified linear units improve restricted boltzmann machines.
\newblock In \emph{ICML}, 2010.

\bibitem[Hendrycks and Gimpel(2016)]{hendrycks2016gaussian}
Dan Hendrycks and Kevin Gimpel.
\newblock Gaussian error linear units {(GELUs)}.
\newblock \emph{arXiv preprint arXiv:1606.08415}, 2016.

\bibitem[Unke et~al.(2021)Unke, Chmiela, Gastegger, Sch{\"u}tt, Sauceda, and
  M{\"u}ller]{unke2021spookynet}
Oliver~T. Unke, Stefan Chmiela, Michael Gastegger, Kristof~T. Sch{\"u}tt,
  Huziel~E. Sauceda, and Klaus-Robert M{\"u}ller.
\newblock {SpookyNet}: Learning force fields with electronic degrees of freedom
  and nonlocal effects.
\newblock \emph{arXiv preprint arXiv:2105.00304}, 2021.

\bibitem[Bernstein(1912)]{bernstein1912demo}
Serge Bernstein.
\newblock Démonstration du théorème de {Weierstrass} fondée sur le calcul
  des probabilités.
\newblock \emph{Comm. Kharkov Math. Soc.}, 13\penalty0 (1):\penalty0 1--2,
  1912.

\bibitem[Hermann et~al.(2020)Hermann, Sch{\"a}tzle, and
  No{\'e}]{hermann2020deep}
Jan Hermann, Zeno Sch{\"a}tzle, and Frank No{\'e}.
\newblock Deep-neural-network solution of the electronic {Schr{\"o}dinger}
  equation.
\newblock \emph{Nat. Chem.}, pages 1--7, 2020.

\bibitem[Neese(2012)]{neese2012orca}
Frank Neese.
\newblock The {ORCA} program system.
\newblock \emph{Wiley Interdisciplinary Reviews: Computational Molecular
  Science}, 2\penalty0 (1):\penalty0 73--78, 2012.

\bibitem[Neese(2018)]{neese2018software}
Frank Neese.
\newblock Software update: the {ORCA} program system, version 4.0.
\newblock \emph{Wiley Interdisciplinary Reviews: Computational Molecular
  Science}, 8\penalty0 (1):\penalty0 e1327, 2018.

\bibitem[Hoffmann(1963)]{hoffmann1963extended}
Roald Hoffmann.
\newblock An extended h{\"u}ckel theory. i. hydrocarbons.
\newblock \emph{The Journal of Chemical Physics}, 39\penalty0 (6):\penalty0
  1397--1412, 1963.

\bibitem[Sch{\"u}tt et~al.(2019)Sch{\"u}tt, Gastegger, Tkatchenko, M{\"u}ller,
  and Maurer]{schutt2019unifying}
Kristof~T. Sch{\"u}tt, Michael Gastegger, Alexandre Tkatchenko, Klaus-Robert
  M{\"u}ller, and Reinhard~J. Maurer.
\newblock Unifying machine learning and quantum chemistry with a deep neural
  network for molecular wavefunctions.
\newblock \emph{Nature Communications}, 10:\penalty0 5024, 2019.

\bibitem[Gastegger et~al.(2020)Gastegger, McSloy, Luya, Sch{\"u}tt, and
  Maurer]{gastegger2020deep}
Michael Gastegger, Adam McSloy, M~Luya, Kristof~T. Sch{\"u}tt, and Reinhard~J.
  Maurer.
\newblock A deep neural network for molecular wave functions in quasi-atomic
  minimal basis representation.
\newblock \emph{The Journal of Chemical Physics}, 153\penalty0 (4):\penalty0
  044123, 2020.

\bibitem[Smith et~al.(2020)Smith, Burns, Simmonett, Parrish, Schieber,
  Galvelis, Kraus, Kruse, Di~Remigio, Alenaizan, et~al.]{smith2020psi4}
Daniel~GA Smith, Lori~A Burns, Andrew~C Simmonett, Robert~M Parrish, Matthew~C
  Schieber, Raimondas Galvelis, Peter Kraus, Holger Kruse, Roberto Di~Remigio,
  Asem Alenaizan, et~al.
\newblock Psi4 1.4: Open-source software for high-throughput quantum chemistry.
\newblock \emph{The Journal of Chemical Physics}, 152\penalty0 (18):\penalty0
  184108, 2020.

\bibitem[Reddi et~al.(2019)Reddi, Kale, and Kumar]{reddi2019convergence}
Sashank~J Reddi, Satyen Kale, and Sanjiv Kumar.
\newblock On the convergence of {Adam} and beyond.
\newblock \emph{arXiv preprint arXiv:1904.09237}, 2019.

\end{thebibliography}
\makeatletter\@input{xxsupplement.tex}\makeatother
\end{document}


\maketitle

\appendix

\section{Detailed Background}
\label{sec:detailed_background}
\subsection{Quantum chemistry}
\label{subsec:quantum_chemistry}
At the center of quantum chemistry methods lies the electronic Schrödinger equation
\begin{equation}
	\hat{H}_\mathrm{el} \Psi_\mathrm{el} = E_\mathrm{el} \Psi_\mathrm{el}\,,
	\label{eq:schrodinger_equation}
\end{equation}
which describes the physical laws underlying the interactions between nuclei and electrons. Here, $\hat{H}_\mathrm{el}$ is the electronic Hamiltonian operator which describes effects due to the kinetic energy of the electrons, the interactions between electrons and nuclei, as well as the inter-electronic interactions. The wavefunction $\Psi_\mathrm{el}$, an eigenfunction of $\hat{H}_\mathrm{el}$, captures the spatial distribution of electrons and the corresponding eigenvalue $E_\mathrm{el}$ represents the electronic energy of the system.

Before the eigenvalue problem can be solved, a suitable functional expression for $\Psi_\mathrm{el}$ has to be found.
A standard approach is to express the wavefunction as a Slater determinant $\Psi_\mathrm{el} = | \psi_1 \ldots \psi_n \rangle$, an anti-symmetric product of molecular orbitals $\psi_i$, which are constructed as linear combinations of atom-centered basis functions $\psi_i = \sum_j C_{ij} \phi_j$.
These atomic orbitals $\phi_j$  are typically taken to be products of radial functions $R_l$ and spherical harmonics $Y^m_l$
\begin{equation}
\phi_i(\mathbf{r}) = R_l(|\mathbf{r}|) Y_{l}^m(\mathbf{r}),
\label{eq:basis}
\end{equation}
where $\mathbf{r}$ denotes the electronic coordinates. Using this ansatz for the wavefunction leads to
\begin{equation}
\mathbf{H}\mathbf{C} = \bm{\epsilon}\mathbf{S}\mathbf{C}\,,
\label{eq:schrodinger_matrix_equation}
\end{equation}
where the Hamiltonian is written as a matrix $\mathbf{H}$ with entries $H_{ij} = \int \phi_i^*(\mathbf{r}) \hat{H}_{\mathrm{el}} \phi_j(\mathbf{r}) d\mathbf{r}$. The overlap matrix $\mathbf{S}$ with entries $S_{ij} = \int \phi_i^*(\mathbf{r})\phi_j(\mathbf{r}) d\mathbf{r}$ has to be introduced and a generalized eigenvalue problem must be solved, because the basis functions $\phi$ are usually not orthonormal. The eigenvectors $\mathbf{C}$ specify the wavefunction $\Psi_\mathrm{el}$ via the coefficients $C_{ij}$ of the molecular orbitals $\psi_i$ and the eigenvalues $\bm{\epsilon}$ are the corresponding orbital energies. Unfortunately, the entries of $\mathbf{H}$ depend on $\mathbf{C}$, because the many-body inter-electronic interactions depend on the positions of all electrons. In other words, $\mathbf{H}$ cannot be determined without knowing $\mathbf{C}$, which in turn cannot be determined without knowing $\mathbf{H}$. To still be able to solve Eq.~\ref{eq:schrodinger_matrix_equation}, approximations have to be introduced.

In the Hartree-Fock formalism, also known as self-consistent field (SCF) method, the Hamiltonian is replaced by the Fock matrix $\mathbf{F} = \mathbf{H}^{\rm core}+\mathbf{G}$, which consists of a one-electron core Hamiltonian $\mathbf{H}^{\rm core}$ (describing the kinetic energy of the electrons and their interaction with nuclei) and a two-electron part $\mathbf{G}$.
Then, starting from an initial guess for $\mathbf{C}$, the matrix $\mathbf{G}$ is computed by approximating the true electron-electron interaction by letting each electron interact with the mean field caused by all other electrons (neglecting correlation effects). The resulting Fock matrix $\mathbf{F}$ is used to solve Eq.~\ref{eq:schrodinger_matrix_equation} (replacing $\mathbf{H}$), leading to updated coefficients $\mathbf{C}$. The two-electron part $\mathbf{G}$ of the Fock matrix is updated using the newly determined coefficients and the procedure is repeated until a self-consistent solution is found. Once a converged solution is found, the total ground state energy $E$ of the chemical system is obtained as
\begin{equation}
E = \sum_{i \in \phi^\mathrm{occ}} \left( \epsilon^\mathrm{core}_i + \epsilon_i \right) + \frac{1}{2}\sum_{I,J}\frac{Z_I Z_J}{\lVert\mathbf{R}_{IJ}\rVert}.
\label{eq:ground_state_en}
\end{equation}
The first sum runs over all occupied (lowest-energy) orbitals, where $\epsilon_i$ are the entries of $\boldsymbol{\epsilon}$ (see Eq.~\ref{eq:schrodinger_matrix_equation}) and $\epsilon^\mathrm{core}_i = \mathrm{diag}(\mathbf{C}^*\mathbf{H}^{\rm core}\mathbf{C})$.
The second term accounts for the classical Coulomb repulsion between nuclei, where $Z_I, Z_J$ are the nuclear charges and $\lVert\mathbf{R}_{IJ}\rVert$ is the distance between two nuclei. 
The forces acting on nuclei can be computed by differentiating the energy in Eq.~\ref{eq:ground_state_en} with respect to the nuclear coordinates, i.e.\ the force acting on nucleus $I$ is given by $-\frac{\partial E}{\partial \mathbf{R}_I}$.

The computational cost of the HF method scales $\mathcal{O}(N^3)$ with the number of basis functions $N$ and is dominated by the iterative procedure and the need to re-evaluate the matrix $\mathbf{F}$ (more precisely, the two-electron part $\mathbf{G}$) whenever the coefficients $\mathbf{C}$ change, which is costly since it involves two-electron integrals. The biggest downside of the HF solution is its low accuracy due to the neglect of electron correlation.
To overcome this limitation, so-called post-HF methods like coupled cluster theory have been developed, using the HF wavefunction as a starting point. However, the improved accuracy of such approaches comes at a significantly higher computational cost (for example $\mathcal{O}(N^7)$), making them prohibitively expensive for large chemical systems.

An efficient alternative to HF and post-HF methods is density functional theory (DFT), where the wavefunction is replaced by the electron density. In this framework, electron correlation can in theory be treated exactly via the so-called exchange-correlation functional.
While DFT scales $\mathcal{O}(N^3)$, the exact form of this functional is unknown and must be approximated with empirical functions, limiting the accuracy of DFT.
Conveniently, DFT in its most frequently used formulation (Kohn-Sham DFT) can be cast in a similar matrix form as Eq.~\ref{eq:schrodinger_matrix_equation}, where the Fock matrix is replaced by the Kohn-Sham matrix $\mathbf{K}$.

\subsection{Group representations and equivariance}
\label{sec:equivariance}
A representation $D$ of a group $G$ is a function from $G$ to square matrices such that for all $g,h \in G$
\begin{equation}
\label{eq:group_representation}
D(g)D(h) =D(gh).
\end{equation}
A function $f:\mathcal{X} \mapsto \mathcal{Y}$, where $\mathcal{X}$ and $\mathcal{Y}$ are vector spaces, is called equivariant with respect to a group $G$ and representations $D^\mathcal{X}$ and $D^\mathcal{Y}$ if for all $g \in G, \mathbf{x}\in\mathcal{X}$, and $\mathbf{y}\in\mathcal{Y}$:
\begin{equation}\label{eq:equivariance}
f(D^\mathcal{X}(g)\mathbf{x}) = D^\mathcal{Y}(g)f(\mathbf{x}).
\end{equation}
When $D^\mathcal{Y}(g)$ is the identity function, the function $f$ is said to be invariant with respect to $G$. Since the composition of two equivariant functions $f_1$ and $f_2$ is also equivariant, any deep neural network that is composed of layers of equivariant functions is itself equivariant.

In the case of the group of 3D rotations, known as $\mathrm{SO}(3)$, any representation $g\in \mathrm{SO}(3)$ can be decomposed as a direct sum of irreducible representations of dimension $2l + 1$, where we call $l$ the degree of the representation. The irreducible linear operators of $\mathrm{SO}(3)$ are known as Wigner-D matrices, with a Wigner-D matrix of degree $l$ being denoted as $\mathbf{D}^{(l)} \in \mathbb{R}^{(2l+1)\times(2l+1)}$.

\subsection{Spherical harmonics}
\label{subsec:spherical_harmonics}
Spherical harmonics $Y^{m}_{l}$ of degree $l=0,\dots,\infty$ and order $m=-l,\dots,l$ form a complete orthonormal basis for functions on the surface of a sphere and are the irreducible representations (irreps) of the 3D rotation group $\mathrm{SO}(3)$. Let $\mathcal{R}(g) \in \mathbb{R}^{3\times 3}$ represent the 3D rotation matrix corresponding to some group element $g \in \mathrm{SO}(3)$ and $\mathbf{D}^{(l)}(g) \in \mathbb{R}^{(2l+1)\times (2l+1)}$ denote the Wigner-D matrix of degree $l$ representing $g$, then for all $g \in \mathrm{SO}(3)$, $\mathbf{r} \in \mathbb{R}^3$:
\begin{equation}
    Y_{l}^{m}(\mathcal{R}(g)\mathbf{r}) = \sum\limits_{m'}\mathbf{D}^{(l)}_{mm'}(g)Y_{l}^{m'}(\mathbf{r}).
\end{equation}
In our implementation we use real-valued spherical harmonics, which for a given degree $l\ge0$ and order $-l\le m \le l$ are defined as
\begin{equation}
\begin{aligned}
Y_l^m(\mathbf{r}) &= \sqrt{\frac{2l+1}{2\pi}}\Pi_l^{\lvert m\rvert}(z)\begin{cases}
\displaystyle
\sum_{p=0}^{\lvert m\rvert}\binom{\lvert m\rvert}{p}x^{p} y^{\lvert m\rvert-p} \sin\left((\lvert m\rvert-p)\frac{\pi}{2}\right)
& m < 0 \\
\displaystyle
\frac{1}{\sqrt{2}}
& m = 0 \\
\displaystyle
\sum_{p=0}^{m}\binom{m}{p}x^{p} y^{m-p} \cos\left((m-p)\frac{\pi}{2}\right)
& m > 0 \\
\end{cases}\,,\\
\Pi_l^m(z) &= \sqrt{\frac{(l-m)!}{(l+m)!}}\sum_{k=0}^{\lfloor(l-m)/2\rfloor}(-1)^k 2^{-l} \binom{l}{k}\binom{2l-2k}{l} \frac{(l-2k)!}{(l-2k-m)!}r^{2k-l}z^{l-2k-m}\,,
\end{aligned}
\label{eq:spherical_harmonics}
\end{equation}
where $x$, $y$, and $z$ are the Cartesian components of vector $\mathbf{r}\in \mathbb{R}^3$ and $r=\lVert\mathbf{r}\rVert$. 

\begin{figure}
	\includegraphics[width=\textwidth]{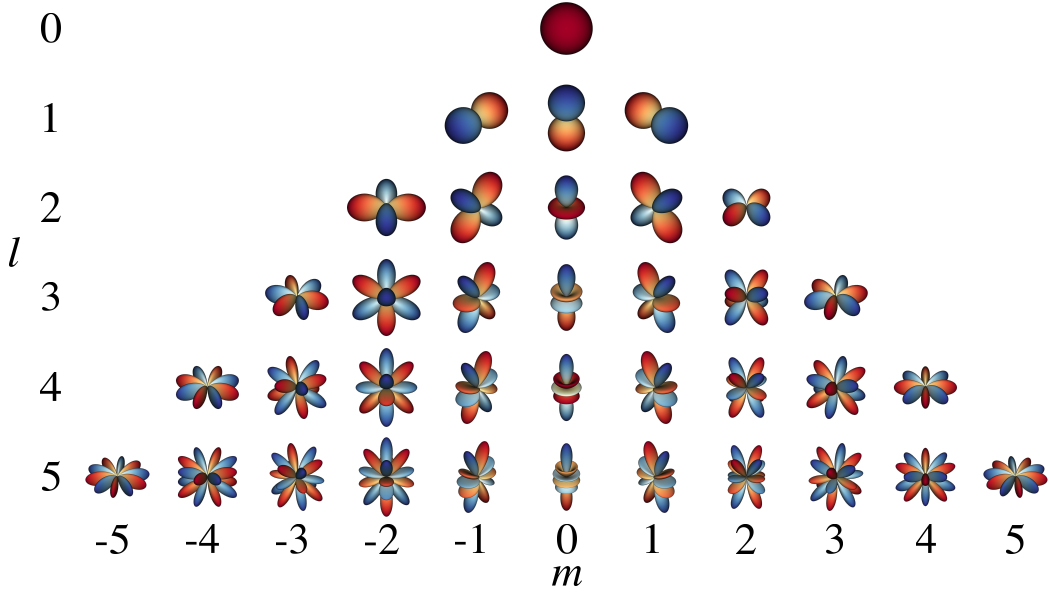}
	\caption{\footnotesize Visualization of spherical harmonics $Y_l^m$ (Eq.~\ref{eq:spherical_harmonics}) of degree $l \in \{0,\dots,5\}$ and order $m \in \{-l,\dots,l\}$ (red and blue indicate positive and negative values, respectively). There are $(L+1)^2$ different possible combinations of $l$ and $m$ for a maximum degree $L$.}
	\label{fig:spherical_harmonics}
\end{figure}
\subsection{Group equivariance in machine learning}
One of the first works to adopt a group theoretic approach to constructing equivariant neural networks was by \citet{cohen2016group}, where equivariance for discrete finite groups is achieved by transforming the convolutional kernel or feature representation according to each group element and aggregating the results. While the method was initially developed for images, it could be extended to other types of data such as vector fields~\citep{marcos2017rotation}. However, the method was limited to applications with discrete groups and a small number of group elements, since the kernels/features need to be explicitly transformed for every group element.

\citet{cohen2016steerable} also explored the possibility of representing convolution kernels as a linear combination of equivariant basis functions, introducing the general concept and demonstrating it for the discrete group of $90^\circ$ rotations. \citet{worrall2017harmonic} used circular harmonics as a basis for the convolutional filters to achieve equivariance to the continuous SO(2) group of 2D rotations. This principle was soon extended to the SO(3) group of 3D rotations by using sphercal harmonics as basis functions \citep{Grisafi2018, cohen2018spherical, thomas2018tensor, weiler20183d, kondor2018clebsch}, which allows applications to 3D structures such as spherical images, voxel data, and point clouds, including atomistic systems. Similarly, \citet{kondor2018covariant} and \citet{hy2018predicting} used group theoretical principles to develop a permutationally equivariant graph neural network to predict properties for atomistic systems.

Most recently, there has been further theoretical research on equivariant neural networks, with ~\citet{cohen2019gauge} extending the principle of equivariance from global symmetries to local gauge transformations, allowing the implementation of a very efficient alternative to spherical CNNs~\citep{cohen2018spherical}, \citet{NEURIPS2019_ea9268cb} providing a proof of the universailty of equivariant graph networks with a single hidden layer, and works such as \citep{NEURIPS2019_b9cfe8b6,finzi2020generalizing} aiming to provide a general framework for the analysis and construction of equivariant networks for a wide range of problems. Very recently, \citet{fuchs2020se} introduced the SE(3)-transformer~\cite{fuchs2020se}, an equivariant generalization of the popular transformer architecture~\citep{vaswani2017attention}.

In this work, we focus on introducing equivariant building blocks mainly used in the context of message-passing neural networks (MPNNs)~\citep{gilmer2017neural}, which are applicable to a variety of graph based problems. From this perspective, our work is most closely related to Cormorant~\citep{anderson2019cormorant} and tensor field networks and their variants~\citep{thomas2018tensor,Batzner2021_e3nn_molecular_dynamics}, which are rotationally equivariant MPNNs using spherical harmonics and SO(3) irreps as equivariant representations.
However, the architectures, feature representations, and operations used in these works have some significant differences to some of the most successful rotationally invariant MPNNs for chemical applications (e.g.\ \citep{schutt2017schnet,unke2019physnet}). When designing our proposed SE(3)-equivariant operations, the aim was to mimic the main building blocks of rotationally invariant MPNNs as closely as possible, such that existing architectures can be ``translated'' to an equivariant framework. Our hope was to benefit from design principles that proved to be successful in previous works, without the need to re-design a successful architecture around the concept of rotational equivariance from scratch. As a result, our proposed PhiSNet architecture (based on the rotationally invariant PhysNet~\citep{unke2019physnet}) has several key differences compared to previous equivariant MPNNs. 

For example, Cormorant does not use explicit activation functions and instead relies on an operation similar to our proposed tensor product contractions as a source of non-linearities, which can cause difficulties during training. While tensor field networks apply activation functions to scalar features (similar to our work), couplings between features of different degrees are only possible in convolution layers, where two (or more) different representations are interacting. In contrast, the operations proposed in this work allow activation functions to affect all degrees of all feature channels through the use of $\mathrm{selfmix}$ layers. 
Our proposed spherical linear layer (which contains a $\mathrm{selfmix}$ operation) acts as a drop-in replacement of linear layers in invariant architectures and allows couplings between all channels and degrees of feature representations.
Additionally, we introduce tensor product expansions of irreducible representations as a general way to construct highly complex second order tensors, while preserving their equivariance properties.

\section{Additional details on SE(3)-equivariant neural network building blocks}
\label{sec:additional_details_network_building_blocks}

\begin{figure}
  \centering
	\includegraphics[width=\textwidth]{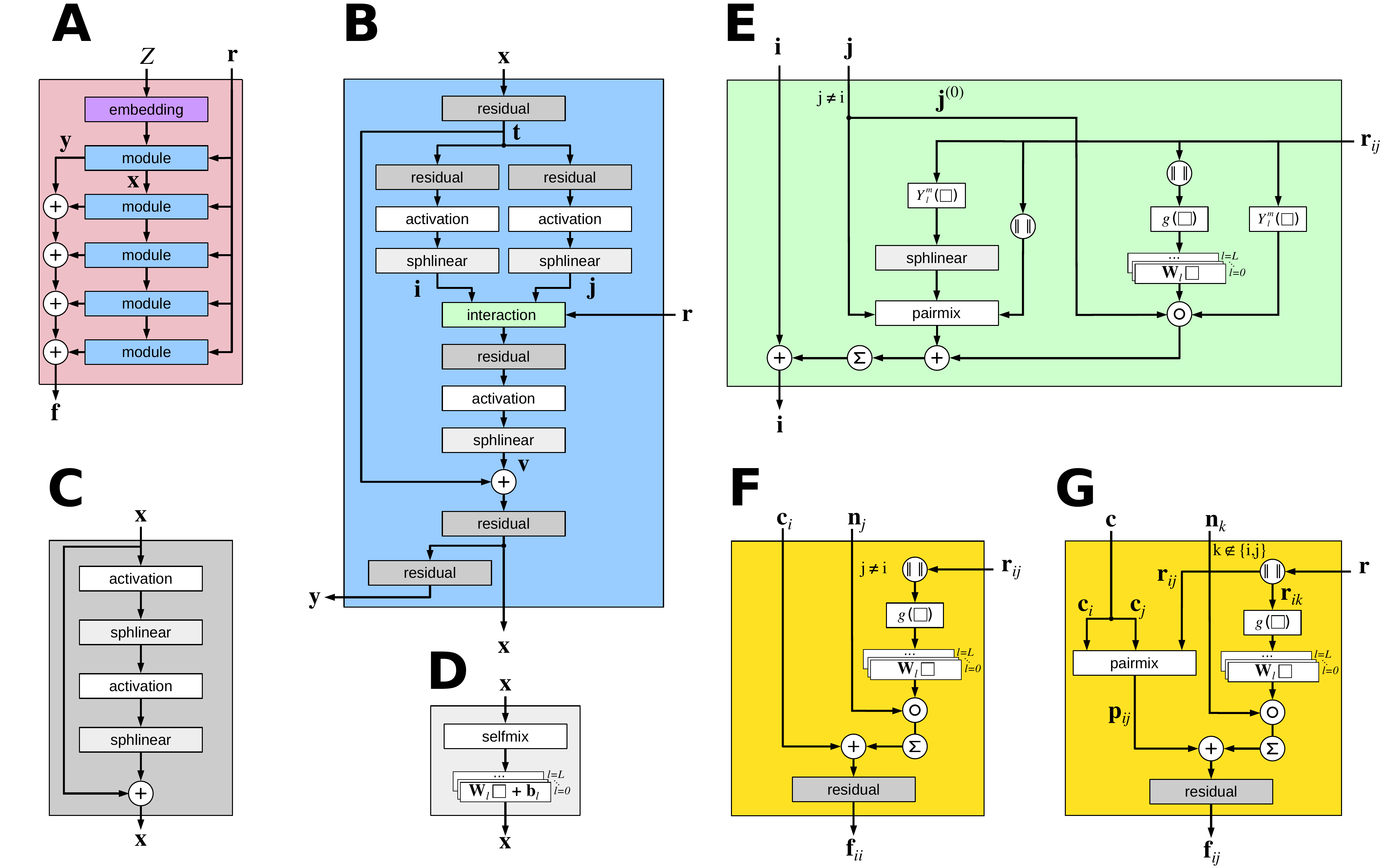}
  \caption{\footnotesize Illustration of PhiSNet components. \textbf{A}: Generation of atomic spherical harmonics features $\mathbf{f}$. The Cartesian coordinates $\{\mathbf{r}_i\}$ of the atoms are used to calculate a spherical harmonics representation of the relative atomic positions. An embedding layer (purple) creates initial atomic features from the nuclear charges $\{Z_i\}$, which are refined through a series of equivariant modular blocks (blue). The outputs $\mathbf{y}$ of each modular block are summed to obtain the final representations $\mathbf{f}$. \textbf{B}: Modular block. Input representations are separated into two branches, which produce separate features for the central $\mathbf{i}$ and neighboring atoms $\mathbf{j}$. These are coupled by an interaction block (green) and added to the original features to produce the $\mathbf{x}$ and $\mathbf{y}$ outputs. \textbf{C}: Residual blocks (Eq.~\ref{Meq:residual_block}) pass input features through two non-linear activations and spherical linear layers and add the result to the unmodified input via a skip-connection. \textbf{D}: Spherical linear layers (Eq.~\ref{Meq:spherical_linear_layer}) are composed of a selfmix layer (Eq.~\ref{Meq:selfmix_layer}) followed by separate linear layers for each spherical harmonic order and are used to re-combine spherical harmonic orders and feature channels. \textbf{E}: The interaction block (Eq.~\ref{Meq:interaction_block}) encodes information about chemical environments by combining features of neighboring atoms with a spherical harmonics based representation of their relative position to a central atom. \textbf{F}: Self-interaction features are generated by updating central atom features with features of neighboring atoms, similar to the interaction block (Eq.~\ref{Meq:self_interaction_feats}). \textbf{G}: Pair-interaction features are obtained by combining the features of a pair of atoms with a pairmix layer and interacting them with the neighboring atoms of the first atom in each pair (Eq.~\ref{Meq:pair_interaction_feats}).} 
\label{fig:nn_architecture}
\end{figure}

\subsection{Properties of the generalized SiLU activation function}
\label{subsec:activation_function}
\begin{figure}[h]
\includegraphics[width=\textwidth]{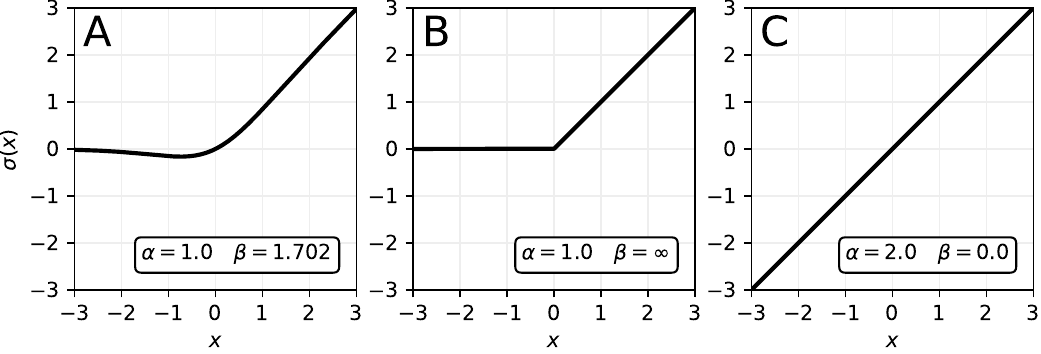}
\caption{\footnotesize Generalized SiLU activation (see Eq.~\ref{Meq:activation_function2}). (A) When the parameters are set to $\alpha=1.0$, $\beta=1.702$, Eq.~\ref{Meq:activation_function2} approximates the GELU function~\citep{hendrycks2016gaussian}. (B) For $\beta \to \infty$, $\sigma(x) \to \alpha\cdot \mathrm{max}(0,x)$, i.e.\ $\sigma(x)$ approaches a (scaled) ReLU activation~\citep{nair2010rectified}. (C) When $\beta$ is zero, Eq.~\ref{Meq:activation_function2} is a linear function with slope $\frac{\alpha}{2}$.}
\label{fig:activation_function}
\end{figure}

Depending on the values of $\alpha$ and $\beta$, the generalized SiLU activation (Eq.~\ref{Meq:activation_function2}) smoothly interpolates between a linear function and the popular ReLU activation~\citep{nair2010rectified} (see Fig.~\ref{fig:activation_function}).  The initial values are chosen as $\alpha = 1.0$ and $\beta=1.702$ such that Eq.~\ref{Meq:activation_function2} approximates the GELU function~\citep{hendrycks2016gaussian}.

\subsection{Exponential Bernstein polynomial basis functions}
\label{subsec:bernstein_polynomials}
\begin{figure}[h]
	\includegraphics[width=\textwidth]{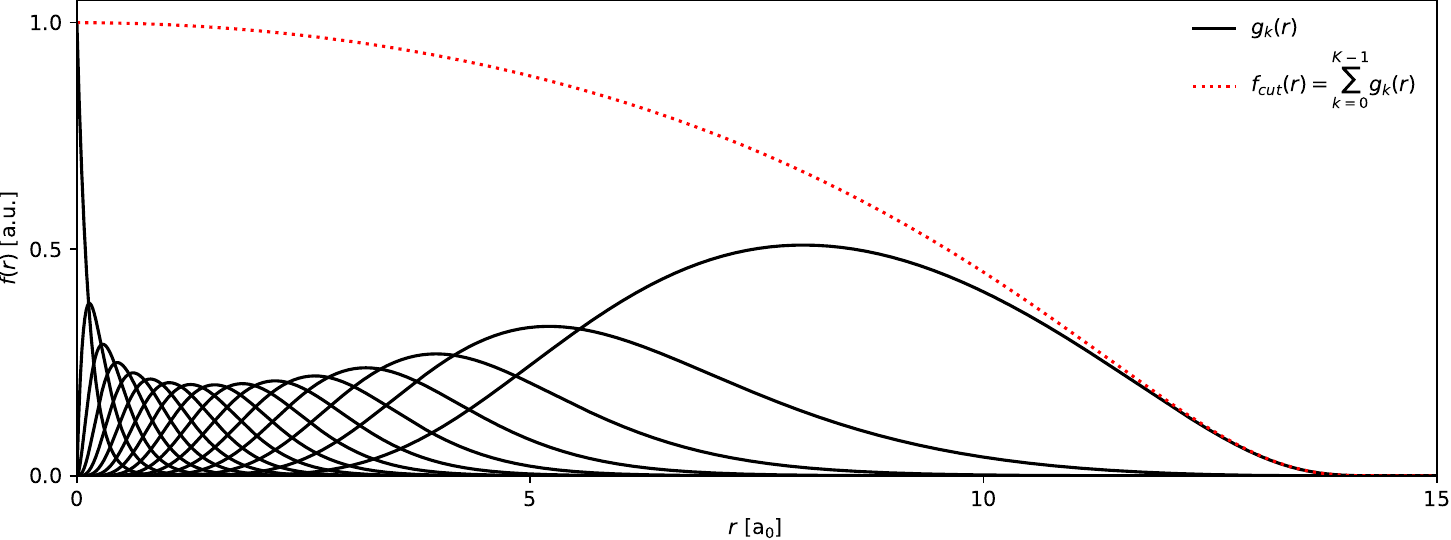}
	\caption{Exponential Bernstein basis functions $g_k(r)$ for $K=16$, $\gamma = 0.5\ \mathrm{a}_0^{-1}$, and $r_{\rm cut}=15\ \mathrm{a}_0$ (see Eq.~\ref{eq:bernstein_basis_function}). Since the Bernstein basis polynomials form a partition of unity, the sum $\sum_k g_k(r)$ is equal to $f_{\rm cut}(r)$.}
	\label{fig:radial_basis_functions}
\end{figure}

All distance-dependant SE(3)-equivariant operations (e.g.\ $\mathrm{pairmix}$ layers) rely on a vector $\mathbf{g} = [g_0(r)\ g_1(r)\ \dots\ g_{K-1}(r)]^{\top}$ representing a basis expansion of the distance via exponential Bernstein radial basis functions~\citep{unke2021spookynet}  given by
\begin{equation}
\begin{aligned}
g_k(r) &= b_{K-1,k}\left(e^{-\gamma r}\right)f_{\rm cut}(r) \\
b_{\nu,n}(x)&=\binom{n}{\nu}x^{\nu}(1-x)^{n-\nu} \qquad \nu=0,\dots,n\\
f_{\rm cut}(r) &= \begin{cases}
\exp\left(-\dfrac{r^2}{(r_{\rm cut}-r)(r_{\rm cut}+r)}\right) & r < r_{\rm cut}\\
0 & r \geq r_{\rm cut}
\end{cases}\,,
\end{aligned}
\label{eq:bernstein_basis_function}
\end{equation}
where $b_{\nu,n}(x)$ are ordinary Bernstein basis polynomials.
For $n\to\infty$, linear combinations of $b_{\nu,n}(x)$ approximate any continuous function on the interval $[0,1]$ uniformly~\citep{bernstein1912demo}. The transformation $x=e^{-\gamma r}$ maps distances $r$ from $[0,\infty]$ to $[0,1]$ and introduces a chemically meaningful inductive bias, i.e.\ wave functions of electrons are known to decay exponentially with increasing distance from a nucleus (a similar mapping is also used in \citep{unke2019physnet,hermann2020deep}). The parameter $\gamma$ is a learnable ``length scale'' and the cutoff function $f_{\rm cut}(r)$ ensures that $g_k(r)$ smoothly goes to zero for $r \geq r_{\rm cut}$ (see Fig.~\ref{fig:radial_basis_functions}). For computational efficiency, all radial basis function expansions across all layers share the same $\gamma$. This way, only one vector $\mathbf{g}(r)$ needs to be computed for each relevant distance $r$.

\section{Additional details on the PhiSNet architecture}
\label{sec:additional_details_phisnet} 

\subsection{Element descriptors used in embeddings}
\label{subsec:embedding}
The embeddings used in the PhiSNet architecture rely on element descriptors $\mathbf{d}_Z$, which encode information about the nuclear charge and the ground state configuration of each element. For example, oxygen ($Z=8$) has the ground state configuration 1s$^2$2s$^2$2p$^4$ and its corresponding descriptor is $\mathbf{v}_8= [8\ 2\ 2\ 4]^{\top}$ (see Table~\ref{tab:species_descriptor} for more examples). In this work, only elements from periods 1 and 2 of the periodic table are considered, i.e.\ 4-dimensional descriptors (with entries for $Z$, 1s, 2s, and 2p electrons) are sufficient. To include elements from higher periods, the descriptors could be extended with additional entries (for example for 3s and 3p electrons for period 3 elements). While the bias term $\mathbf{b}_Z$ in Eq.~\ref{Meq:embedding_layer} offers sufficient degrees of freedom to learn arbitrary embeddings for different elements, including the term $\mathbf{W}\mathbf{d}_Z$ provides an inductive bias that takes into account the quantum chemical structure of different elements. Similar element descriptors are also used in \citep{unke2021spookynet}.
\begin{table}[h]
  \centering
	\begin{tabular}{c c c c c c} 
	    \toprule
		element & Z & 1s & 2s & 2p & $\mathbf{d}_Z^{\top}$ \\
        \midrule
		H & 1 & 1 & 0 & 0 & $[1\ 1\ 0\ 0]$\\
		C & 6 & 2 & 2 & 2 & $[6\ 2\ 2\ 2]$\\
		N & 7 & 2 & 2 & 3 & $[7\ 2\ 2\ 3]$\\
		O & 8 & 2 & 2 & 4 & $[8\ 2\ 2\ 4]$\\
		\bottomrule
	\end{tabular}
	\caption{\footnotesize Examples of element descriptors $\mathbf{d}_Z$.}
	\label{tab:species_descriptor}
\end{table}

\subsection{Details on the block-wise Hamiltonian matrix prediction}

\begin{figure}[h]
  \centering
	\includegraphics[width=\textwidth]{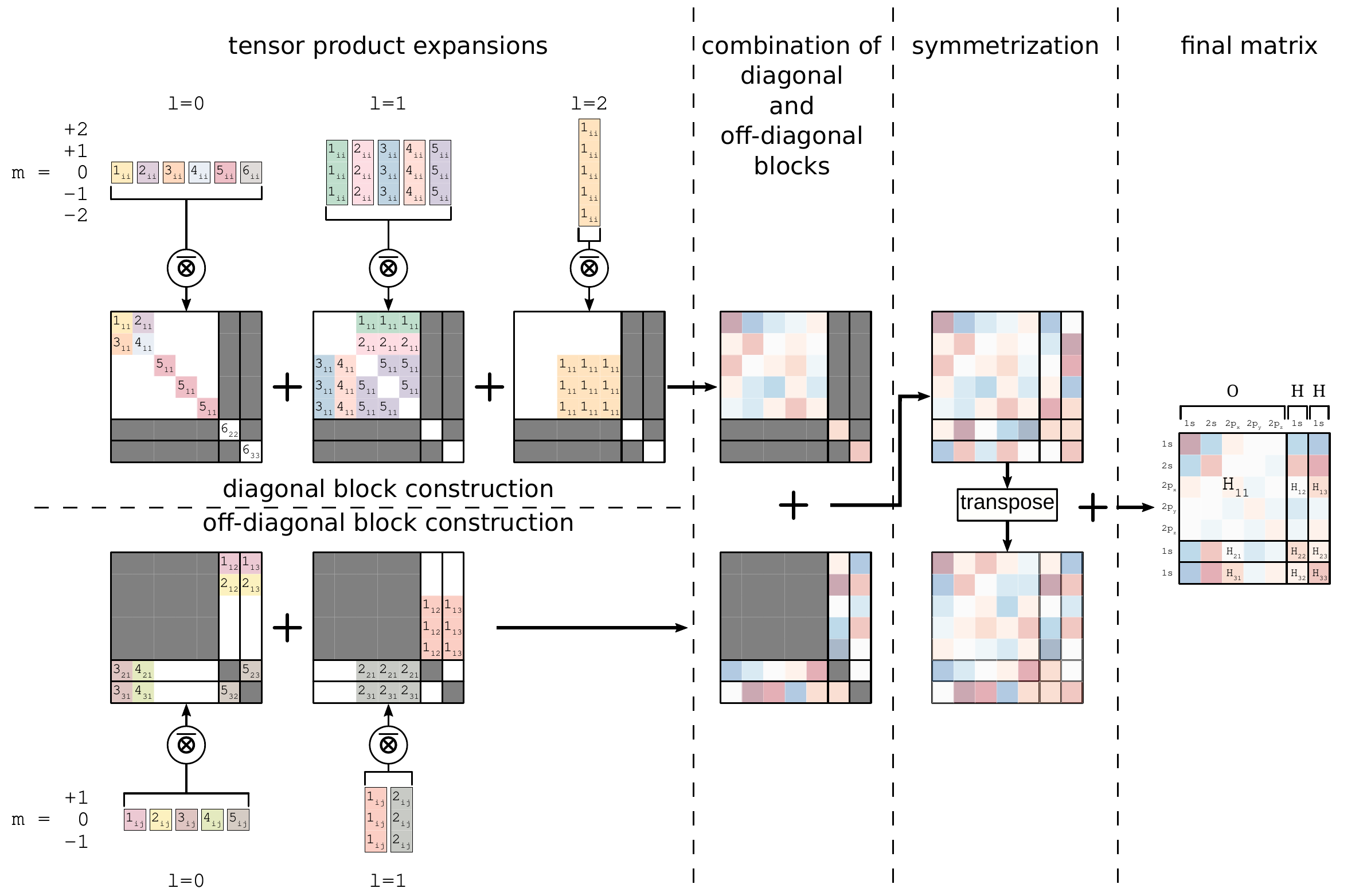}
   \caption{\footnotesize An illustration for the construction of the Hamiltonian matrix for a water molecule using a minimal basis set, showing how irreps of different degrees are transformed via tensor product expansions to reconstruct the Hamiltonian matrix block-by-block. The irreps are represented by colored squares labelled $c_{ii}$ or  $c_{ij}$, where $c$ is the channel index (see Tab~\ref{tab:h2o_irrep_breakdown}) and the subscript specifies atom indices. Different orders of an irrep are represented by individual squares of the same color. 
   Expanded irreps are placed at the appropriate position in the matrix depending on the orbital-pair interaction that the block represents. The individual blocks are then added together to form the diagonal and off-diagonal parts of the matrix, after which the two parts are combined and the result is added to its transpose in order to ensure the final result is a symmetric matrix.}
	\label{fig:hamiltonian_composition}
\end{figure}

To illustrate the process of the block-wise Hamiltonian matrix construction for a concrete example, let us consider a water molecule using a minimal basis set, i.e.\ a single basis function is used to represent each atomic orbital. Water consists of two H atoms ($Z=1$) with a 1s ($l=0$) orbital, and one O  atom ($Z=8$) with a 1s ($l=0$), a 2s ($l=0$), and a 2p ($l=1$) orbital. An orbital with degree $l$ has $2l+1$ distinct orders $m$, so the complete Hamiltonian is a $7\times7$ matrix. In general, to represent the $(2l_1+1)\times(2l_2+1)$ matrix block that corresponds to the interaction between two orbitals of degrees $l_1$ and $l_2$, we need irreps for all degrees $l_3 \in \{|l_1-l_2|,...,l_1+l_2\}$. In the case of water with a minimal basis set, there are three main cases to consider:
\begin{enumerate}
    \item Interaction between two s-orbitals ($l_1=0, l_2=0$), i.e.\ $\mathbb{1}\otimes\mathbb{1}=\mathbb{1}$ ($1\times 1$ matrix block, an irrep of degree $l_3=0$ is needed) 
    \item Interaction between an s- and a p-orbital ($l_1=0, l_2=1$), i.e.\ $\mathbb{1}\otimes\mathbb{3}= \mathbb{3}$ ($1\times 3$ matrix block, an irrep of degree $l_3=1$ is needed)
    \item Interaction between two p-orbitals ($l_1=1, l_2=1$), i.e.\ $\mathbb{3}\otimes\mathbb{3}= \mathbb{1}\oplus\mathbb{3}\oplus\mathbb{5}$ ($3\times 3$ matrix block, irreps of degrees $l_3=0,1,2$ are needed)
\end{enumerate}
Now, unique channel indices have to be assigned to collect irreps of all degrees for the interaction between individual orbitals from the self-interaction and pair-interaction features $\mathbf{f}_{ii}$ and $\mathbf{f}_{ij}$. The indices are stored in the corresponding index sets $I^{\rm self}$ and $I^{\rm pair}$ and the total number of feature channels needs to be chosen large enough (such that all assigned indices are valid). To construct the diagonal blocks, 6 irreps of degree 0, 5 irreps of degree 1, and 1 irrep of degree 2 are necessary, whereas for the off-diagonal blocks, 5 irreps of degree 0, and 2 irreps of degree 1 are required (see Table~\ref{tab:h2o_irrep_breakdown} for a complete breakdown). Fig.~\ref{fig:hamiltonian_composition} illustrates how the irreps are transformed to matrix blocks via tensor product expansions (Eq.~\ref{Meq:tensor_product_expansion}) and accumulated to construct the complete Hamiltonian matrix.

\begin{table}[h]
    \centering
    \begin{tabular}{c c c c}
    \toprule
    \multicolumn{4}{c}{\bf irreps for diagonal blocks}\\
    $I^{\rm self}(Z,n,m,L)$ & orbital interaction & degree & index\\
    \midrule
    $I^{\rm self}(8,1,1,0)$ & O:1s $\times$ O:1s $\rightarrow \mathbb{1}$ & 0 & 1  \\
    $I^{\rm self}(8,1,2,0)$  & O:1s $\times$ O:2s $\rightarrow \mathbb{1}$ & 0 & 2\\
    $I^{\rm self}(8,2,1,0)$  & O:2s $\times$ O:1s $\rightarrow \mathbb{1}$ & 0 & 3\\
    $I^{\rm self}(8,2,2,0)$  & O:2s $\times$ O:2s $\rightarrow \mathbb{1}$ & 0 & 4\\
    $I^{\rm self}(8,3,3,0)$  & O:2p $\times$ O:2p $\rightarrow \mathbb{1}$ & 0 & 5 \\
    $I^{\rm self}(1,1,1,0)$  & H:1s $\times$ H:1s $\rightarrow \mathbb{1}$ & 0 & 6\\
    $I^{\rm self}(8,1,3,1)$  & O:1s $\times$ O:2p $\rightarrow \mathbb{3}$ & 1 & 1\\
    $I^{\rm self}(8,2,3,1)$  & O:2s $\times$ O:2p $\rightarrow \mathbb{3}$ & 1 & 2\\
    $I^{\rm self}(8,3,1,1)$  & O:2p $\times$ O:1s $\rightarrow \mathbb{3}$ & 1 & 3\\
    $I^{\rm self}(8,3,2,1)$  & O:2p $\times$ O:2s $\rightarrow \mathbb{3}$ & 1 & 4\\
    $I^{\rm self}(8,3,3,1)$  & O:2p $\times$ O:2p $\rightarrow \mathbb{3}$ & 1 & 5\\
    $I^{\rm self}(8,3,3,2)$  & O:2p $\times$ O:2p $\rightarrow \mathbb{5}$ & 2 & 1\\
    \midrule
    \multicolumn{4}{c}{\bf irreps for off-diagonal blocks}\\
    $I^{\rm pair}(Z_1,Z_2,n_1,n_2,L)$ & orbital interaction & degree & index\\
    \midrule
    $I^{\rm pair}(8,1,1,1,0)$ & O:1s $\times$ H:1s $\rightarrow \mathbb{1}$ & 0 & 1\\
    $I^{\rm pair}(8,1,2,1,0)$ & O:2s $\times$ H:1s $\rightarrow \mathbb{1}$ & 0 & 2\\
    $I^{\rm pair}(1,8,1,1,0)$ & H:1s $\times$ O:1s $\rightarrow \mathbb{1}$ & 0 & 3\\
    $I^{\rm pair}(1,8,1,2,0)$ & H:1s $\times$ O:2s $\rightarrow \mathbb{1}$ & 0 & 4\\
    $I^{\rm pair}(1,1,1,1,0)$ & H:1s $\times$ H:1s $\rightarrow \mathbb{1}$ & 0 & 5\\
    $I^{\rm pair}(8,1,3,1,0)$ & O:2p $\times$ H:1s $\rightarrow \mathbb{3}$ & 1 & 1\\
    $I^{\rm pair}(1,8,1,3,0)$ & H:1s  $\times$ O:2p $\rightarrow \mathbb{3}$ & 1 & 2\\
    \bottomrule
    \end{tabular}
    \caption{\footnotesize Breakdown of all irreps necessary to construct the Hamiltonian matrix for a water molecule using a minimal basis set. Here, we label the 1s orbital of H with the number 1, and the 1s, 2s, and 2p orbitals of O with the numbers 1, 2, and 3. Note that channel indices can be chosen arbitrarily, as long as indices for irreps of the same degree are unique and the assignment is consistent.}
    \label{tab:h2o_irrep_breakdown}
\end{table}

\subsection{Overlap matrix prediction}
\label{subsec:overlap_matrix}
Since the overlap matrix $\mathbf{S}$ represents the overlap integral of atomic orbital basis functions, it is not affected by many-body effects and its entries depend only on the types of interacting elements and their relative orientation and distance. Thus, simplified feature representations can be used to construct it. Self-interaction features for the overlap matrix are obtained by passing the initial atomic feature representations $\mathbf{x}$ (produced by the embedding layer from the nuclear charges $Z$) through a spherical linear layer and a residual block: 
\begin{equation}
\begin{aligned}
  \mathbf{u}_{ii}^{(l)} &= \mathrm{sphlinear}_{L \to L,F \to F}(\mathbf{x}_i)^{(l)}\,,\\
  \mathbf{s}_{ii} &= \mathrm{residual}(\mathbf{u}_{ii})\,.
\end{aligned}
\label{eq:self_interaction_overlap}
\end{equation}
Similarly, pair-interaction features are generated according to:
\begin{equation}
\begin{aligned}
  \mathbf{u}_{ij}^{(l)} &= \mathrm{pairmix}_{L,L\to L}(\mathbf{x}_i,\mathrm{sphlinear}_{L\to L,1 \to F}\left(\mathbf{Y}(\mathbf{r}_{ij})\right),\lVert \mathbf{r}_{ij}\rVert)^{(l)}\,,\\
  \mathbf{s}_{ij} &= \mathrm{residual}(\mathbf{u}_{ij})\,.
\end{aligned}
\label{eq:pair_interaction_overlap}
\end{equation}
The overlap matrix is then assembled from the features $\mathbf{s}_{ii}$ and $\mathbf{s}_{ij}$ (insteaf of $\mathbf{f}_{ii}$ and $\mathbf{f}_{ij}$) analogously to other Hamiltonian matrices.

\section{Ablation studies}
\label{sec:ablation_studies}
To explore the impact of different possible architectural simplifications on the prediction accuracy of PhiSNet, the following ablated variants are considered:
\begin{itemize}
    \item \textbf{shared matrix features}: The same features are used for central and neighboring atoms for the Hamiltonian matrix prediction. In other words, the parameters of $\mathrm{residual}$ in Eq.~\ref{Meq:center_neighbor_features} are shared between the computation of $\mathbf{c}$ and  $\mathbf{n}$, i.e.\  $\mathbf{c}=\mathbf{n}$.
    
    \item \textbf{shared interaction features}: The same features are used for central and neighboring atoms in interaction blocks (Eq.~\ref{Meq:interaction_block}), i.e.\ $\mathbf{c}=\mathbf{n}$. In other words, the parameters of $\mathrm{sphlinear}$ and $\mathrm{residual}$ operations are shared between the computation of the $\mathbf{i}$ and $\mathbf{j}$ features in Eq.~\ref{Meq:neural_network_module}.
    
    \item \textbf{no selfmix}: All $\mathrm{selfmix}$ (Eq.~\ref{Meq:selfmix_layer}) operations are removed from spherical linear layers (Eq.~\ref{Meq:spherical_linear_layer}) when $L_{\rm in} = L_{\rm out}$, i.e.\ 
    \begin{equation*}
    \mathrm{sphlinear}_{L \to L, F_{\rm in} \to F_{\rm out}}(\mathbf{x}) =
    \mathrm{linear}_{F_{\rm in} \to F_{\rm out}}\left(\mathbf{x}\right)\,.
    \end{equation*}

    \item \textbf{simple pair features}: No interaction with neighboring atoms is used for constructing self-interaction ($\mathbf{f}_{ii}$, Eq.~\ref{Meq:self_interaction_feats}) and pair-interaction ($\mathbf{f}_{ij}$, Eq.~\ref{Meq:pair_interaction_feats}) features, i.e.\ they are computed as
    \begin{equation*}
        \mathbf{f}_{ii}^{(l)} = \mathrm{residual}\left(\mathbf{c}_i^{(l)}\right)\,,
    \end{equation*}
    \begin{equation*}
        \mathbf{f}_{ij}^{(l)} = \mathrm{residual}\left(\mathrm{pairmix}(\mathbf{c}_i, \mathbf{c}_j, \|\mathbf{r}_{ij}\|)^{(l)}\right)\,.
    \end{equation*}
\end{itemize}

The results of the ablation studies are shown in Table~\ref{tab:ablation_studies}. We find that all considered modifications significantly reduce the prediction accuracy of PhiSNet. With exception of the ``no selfmix'' variant, any improvements in computational efficiency obtained by simplifying the architecture are negligible ($<$1\%) and do not justify the observed reduction in accuracy. The ``no selfmix'' variant, however, although between 14-77\% less accurate than the baseline model, can be a valid alternative for some applications: We find that inference speed and training times are reduced by roughly 50\% for this variant.

\begin{table}[h]
  \caption{Prediction errors (in units of $10^{-6}~\mathrm{E}_{\rm h}$) for Kohn-Sham matrices of ablated variants compared with the baseline architecture. Note that models are not trained to convergence, but for a fixed number of training steps (40k steps for models trained on water, 30k steps for models trained on ethanol).}
 \centering
\begin{tabularx}{0.8\textwidth}{c *{5}{>{\centering\arraybackslash}X}}
\toprule
\multirow{2}{*}{dataset} & \multirow{2}{*}{baseline} & shared & shared & no & simple\\
& & mat. feat. & int. feat. & selfmix & pair feat.\\
\midrule
Water & 19.18 & 24.93 & 29.57 & 33.88 & 88.21 \\
Ethanol &  77.29 & 81.70 & 89.47 & 87.85 & 121.63\\
\bottomrule
\end{tabularx}
\label{tab:ablation_studies}
\end{table}

\section{Speedup with respect to \textit{ab initio} calculations}
\label{sec:speedup_wrt_dft}
\subsection{Model inference times compared to \textit{ab initio} calculations}
\label{subsec:inference_times}
To evaluate the speedup PhiSNet provides compared to \textit{ab initio} calculations, we compare model inference times to DFT calculations at the PBE/def2-SVP level of theory with the ORCA 4.0.1 software~\cite{neese2012orca,neese2018software}. PhiSNet was evaluated on an NVIDIA A100 GPU with a batch size of 64, whereas DFT calculations were performed on Intel Xeon E5-2690 CPUs. The average wall-clock times for evaluating a single structure are summarized in Tab.~\ref{tab:inference_timings}.
\begin{table}[h]
    \centering
    \begin{tabular}{l r r r}
    \toprule
        molecule & DFT & PhiSNet & speedup \\
        \midrule
        ethanol & 21.657 s & 0.027 s & $\sim802\times$ \\
        malondialdehyde & 38.923 s & 0.029 s & $\sim1342\times$ \\
        uracil & 86.675 s & 0.050 s & $\sim1734\times$ \\
        aspirin & 343.615 s & 0.155 s & $\sim2217\times$\\
        \bottomrule
    \end{tabular}
    \caption{Average wall-clock time for evaluating a single structure with DFT and PhiSNet for different molecules.}
    \label{tab:inference_timings}
\end{table}

\subsection{Speeding up \textit{ab initio} calculations using predicted wavefunctions as initial guess}
\label{subsec:initial_guess}
Although PhiSNet achieves low prediction errors with respect to quantities calculated from \textit{ab initio} calculations, even small errors might be unacceptable for applications that require exact solutions. Even then, PhiSNet can speedup quantum chemistry by providing the predicted wavefunctions as initial guess (typically, the guess is obtained from a semi-empirical method such as extended H\"uckel theory~\cite{hoffmann1963extended} or from using model potentials). Tab.~\ref{tab:dft_reduction} summarizes the reduction in iterations until convergence and total wall-clock time for different SCF algorithms when using wavefunctions predicted by PhiSNet instead of the default guess. Note that the quality of guess wavefunctions provided by PhiSNet also makes it possible to use quadratically converging SCF algorithms like SOSCF and NRSCF instead of the standard DIIS approach. All DFT calculations were performed with the ORCA 4.0.1 software~\cite{neese2012orca,neese2018software}.

\begin{table}[h]
    \centering
    \begin{tabular}{ccccccc}
    \toprule
    \bf  molecule & \multicolumn{3}{c}{\bf \# iterations until convergence} & \multicolumn{3}{c}{\bf total wall-clock time} \\
    SCF algorithm & DIIS & SOSCF & NRSCF & DIIS & SOSCF &  NRSCF \\        
    \midrule
    \bf ethanol & 47.1\% & 44.6\% & 68.6\% & 31.5\% & 30.4\% & 36.5\% \\
    \bf malondialdehyde & 51.3\% & 42.4\% & 71.5\% & 40.5\% & 34.3\% & 40.0\% \\
    \bf uracil & 47.3\% & 40.4\% & 67.1\% & 38.1\% & 32.9\% & 38.0\% \\
    \bf aspirin & 33.4\% & 29.5\% & 55.6\% & 29.1\% & 24.8\% & 24.5\% \\
    \bottomrule
    \end{tabular}
    \caption{Average reduction in the number of iterations until convergence and total wall-clock time for different molecules and SCF algorithms when starting from the default versus PhiSNet-predicted guess wavefunction.}
    \label{tab:dft_reduction}
\end{table}


\section{Datasets}
\label{sec:datasets}

The DFT datasets (containing energies, forces, Kohn-Sham matrices and overlap matrices) for water, ethanol, malondialdehyde and uracil were taken from \citep{schutt2019unifying} and are also available on \href{http://quantum-machine.org/datasets/}{http://quantum-machine.org/datasets/}.
The corresponding data for the aspirin molecule was obtained from \citep{gastegger2020deep}.
With the exception of water, these datasets are based on subsets of structures drawn from the MD17 dataset. Water structures were sampled using a classical force field.
All quantities in these datasets have been computed at the PBE/def2-SVP level of theory, where PBE denotes the computational method (in this case density functional) and def2-SVP the atomic basis set used for expanding the wavefunction.
The water datasets used in the transfer learning experiments use the same molecular structures as the DFT water data.
Based on these structures energies, forces, as well as Fock, overlap and core Hamiltonian matrices (the latter three only in the case of HF/cc-pVDZ) were computed at the HF/cc-pVDZ and CCSD(T)/cc-pVTZ levels of theory using the PSI4 code package~\citep{smith2020psi4}.
Further details, such as data set size, are provided in Table~\ref{tab:datasets}.

\begin{table}[h]
    \caption{Datasets used in this work. The level of theory refers to the combination of method and basis set used to calculate molecular properties. $N_\mathrm{mol}$ denotes the number of molecular structures in each dataset, while $N_\mathrm{atoms}$ is the number of atoms.
    $N_\mathrm{basis}$ is the total number of atomic basis functions used to express the wavefunction of a molecule. All matrix properties (Fock matrix, core Hamiltonian, Kohn-Sham matrix and overlap matrix) are of the dimension $N_\mathrm{basis} \times N_\mathrm{basis}$.  We omit this value for the last row, where no matrix quantities have been computed.}
    \label{tab:datasets}
    \centering
    \begin{tabular}{llrrrl}
    \toprule
        Dataset & Level of theory & $N_\mathrm{mol}$ &  $N_\mathrm{atoms}$ & $N_\mathrm{basis}$ & Source\\
    \midrule
        Water           & PBE/def2-SVP    &   4\,999 &  3 &     24 & \citep{schutt2019unifying}\\
        Ethanol         & PBE/def2-SVP    &  30\,000 &  9 &     72 & \citep{schutt2019unifying}\\
        Malondialdehyde & PBE/def2-SVP    &  26\,978 &  9 &     90 & \citep{schutt2019unifying}\\
        Uracil          & PBE/def2-SVP    &  30\,000 & 12 &    132 & \citep{schutt2019unifying}\\
        Aspirin         & PBE/def2-SVP    &  30\,000 & 21 &    222 & \citep{gastegger2020deep}\\
        Water           & HF/cc-pVDZ      &   4\,999 &  3 &     24 & this work\\
        Water           & CCSD(T)/cc-pVTZ &   4\,999 &  3 &      - & this work\\
    \bottomrule
    \end{tabular}
\end{table}

\section{Training procedure and hyperparameters}
\label{sec:training_procedure}
All models in this work use $F=128$ feature channels with a maximum degree $L_{\rm max} = 4$, and $K=128$ basis functions with a cutoff radius of $r_{\rm cut}=15~a_0$ (see Eq.~\ref{eq:bernstein_basis_function}), resulting in roughly 17M parameters (for comparison, SchNOrb~\citep{schutt2019unifying} uses 93M parameters). The parameters were optimized with AMSGrad~\citep{reddi2019convergence} using an initial learning rate of $10^{-3}$, other hyperparameters of the optimizer were set to the recommended defaults. The performance on the validation set was evaluated every 1k training steps and the learning rate decayed by a factor of 0.5 if the validation loss did not decrease for 10 consecutive evaluations. Training was stopped once the learning rate was smaller than $10^{-5}$ and the best-performing model (lowest validation loss) was selected. All models were trained on a single NVIDIA Titan Xp GPU. 

\subsection{DFT datasets}
Models for DFT datasets were trained by minimizing the loss function
\begin{equation}
\mathcal{L} = \frac{1}{N_{\rm batch}} \sum_{b=1}^{N_{\rm batch}}  \left( \lVert \mathbf{K}_b^{\mathrm{ref}}-\mathbf{K}_b\rVert_F^2 + \lVert \mathbf{S}_b^{\mathrm{ref}}-\mathbf{S}_b\rVert_F^2 \right)\,,
\label{eq:dft_loss}
\end{equation}
where $\mathbf{K}_b$ and $\mathbf{S}_b$ are the predicted Kohn-Sham and overlap matrices for structure $b$ in the batch, 
$\mathbf{K}_b^{\mathrm{ref}}$ and $\mathbf{S}_b^{\mathrm{ref}}$ denote the corresponding reference matrices, and $\lVert \cdot \rVert_F^2$ is the squared Frobenius norm. The batch, training set, validation set, and test set sizes used for the different datasets are given in Tab.~\ref{tab:dftsettings}.

\begin{table}[h]
    \caption{Batch, training set, validation set, and test set sizes for models trained on the DFT datasets. Note that batch sizes between different molecules were not tuned for performance. All variations are due to different sized Hamiltonian matrices and the memory constraints of the training hardware (e.g.\ water consists of 3 atoms and Hamiltonian matrices are $24 \times 24$, whereas aspirin consists of 21 atoms and Hamiltonian matrices are $222 \times 222$).}
    \label{tab:dftsettings}
    \centering
    \begin{tabular}{llrrr}
    \toprule
        Dataset & $N_{\rm batch}$ & $N_{\rm train}$  &  $N_{\rm valid}$  & $N_{\rm test}$ \\
    \midrule
        Water  & 50 & 500 & 500 & 3\,999 \\
        Ethanol  & 10 & 25\,000 & 500 & 4\,500 \\
        Malondialdehyde  & 10 & 25\,000 & 500 & 1\,478 \\
        Uracil  & 5 & 25\,000 & 500 & 4\,500 \\
        Aspirin  & 2 & 25\,000 & 500 & 4\,500 \\
    \bottomrule
    \end{tabular}
\end{table}

\subsection{Transfer learning from HF to CCSD(T)}
\label{subsec:transfer_learning}
For the transfer learning experiment, the model is trained on the same structures with data computed both at the HF/cc-pVDZ and CCSD(T)/cc-pVTZ levels of theory (see last two rows of Tab.~\ref{tab:datasets}). We use a batch size of $N_{\rm batch} = 25$, and training, validation, and test sets of sizes $N_{\rm train} = 500$, $N_{\rm valid} = 500$, and $N_{\rm test}=3\,999$. 

First, the model is trained by minimizing the loss function 
\begin{equation}
\mathcal{L}_1 = \frac{1}{N_{\rm batch}} \sum_{b=1}^{N_{\rm batch}} \left( \lVert \mathbf{F}_b^{\mathrm{ref}}-\mathbf{F}_b\rVert_F^2 + \lVert + \lVert \mathbf{H}_b^{\mathrm{core,ref}}-\mathbf{H}_b^{\mathrm{core}}\rVert_F^2 + \lVert \mathbf{S}_b^{\mathrm{ref}}-\mathbf{S}_b\rVert_F^2 \right)\,,
\label{eq:hf_matrix_loss}
\end{equation}
where $\mathbf{F}_b$, $\mathbf{H}_b^{\rm core}$,  $\mathbf{S}_b$ are the predicted Fock, core Hamiltonian, and overlap matrices for structure $b$ in the batch, 
and $\mathbf{F}_b^{\rm ref}$, $\mathbf{H}_b^{\rm core,ref}$,  $\mathbf{S}_b^{\rm ref}$ denote the corresponding HF-level reference matrices. Although a model trained in this way reaches matrix prediction errors that are comparable with the results obtained for DFT datasets (see Tab.~\ref{Mtab:comparison_with_schnorb}), we observe that errors for energies and forces derived from the predicted matrices (using Eq.~\ref{eq:ground_state_en}) are much larger. We speculate that this is due to the fact that all matrix elements are weighed equally when computing $ \mathcal{L}_1$, although some elements will have a much greater effect on the energy and forces than others. For this reason, we re-train our model on the HF data using a new loss function $\mathcal{L} = \mathcal{L}_1 + \mathcal{L}_2$ with
\begin{equation}
\mathcal{L}_2 = \frac{1}{N_{\rm batch}}  \sum_{b=1}^{N_{\rm batch}}\left( (E_b^{\rm{ref}}-E_b)^2 + \frac{1}{N} \sum_{i=1}^{N} \left\lVert-\frac{\partial E_b}{ \partial \mathbf{R}_{b,i}}-\mathbf{f}_{b,i}^{\rm{ref}}\right\rVert^2\right) \,,
\label{eq:hf_energy_force_loss}
\end{equation}
which incorporates information from HF-level energy and forces directly. Here, $E_b$ is the predicted energy for structure $b$ in the batch, $E_b^{\rm ref}$ the corresponding HF reference, $\mathbf{f}_{b,i}^{\rm ref}$ the reference force acting on atom $i$ in structure $b$ and $\mathbf{R}_{b,i}$ its Cartesian coordinates (in total, each structure contains $N$ atoms). Force predictions are obtained using automatic differentiation. After re-training with the modified loss, energy and force predictions are improved substantially, although the prediction of matrix elements deteriorates slightly (see Tab.~\ref{tab:hf_results}).
\begin{table}[h]
	\begin{tabular}{l c c c c c}
		\toprule
         \multirow{2}{*}{\bf model} & $\mathbf{F}$ & $\mathbf{S}$ & $\mathbf{H}_{\rm core}$ & energy & forces \\
		  & [$10^{-6}~\mathrm{E}_{\rm h}$] & [$10^{-6}$] & [$10^{-6}~\mathrm{E}_{\rm h}$] & [$10^{-3}~\mathrm{E}_{\rm h}$] & [$10^{-3}~\mathrm{E}_{\rm h}/a_0$] \\
		
		\midrule
		
trained on  $\mathcal{L}_1$  & 24.77 & 1.06 & 39.90  & 6.51 & 57.10 \\
 re-trained on $\mathcal{L}_1+\mathcal{L}_2$  & 46.94 & 5.32  & 58.00 & 0.117 & 2.289\\
				
		\bottomrule
	\end{tabular}
	\caption{Comparison of mean absolute prediction errors for models trained on HF/cc-pVDZ reference data.}
	\label{tab:hf_results}
\end{table}
We also experimented with training a model directly on $\mathcal{L} = \mathcal{L}_1 + \mathcal{L}_2$, but found that this does not work: The matrices predicted at the beginning of training lead to numerical instabilities when solving the generalized eigenvalue problem (Eq.~\ref{eq:schrodinger_matrix_equation}), i.e.\ a warm-up-phase using the pure matrix loss $\mathcal{L}_1$ is necessary for convergence.

Finally, the re-trained model is fine-tuned using the loss function
\begin{equation}
\mathcal{L}_{\rm TL} = \frac{1}{N_{\rm batch}} \sum_{b=1}^{N_{\rm batch}} \left( \frac{1}{N} \sum_{i=1}^{N} \left\lVert-\frac{\partial E_b}{ \partial \mathbf{R}_{b,i}}-\mathbf{f}_{b,i}^{\rm{ref,CC}}\right\rVert^2  + \lVert \mathbf{H}_b^{\mathrm{core,ref}}-\mathbf{H}_b^{\mathrm{core}}\rVert_F^2 + \lVert \mathbf{S}_b^{\mathrm{ref}}-\mathbf{S}_b\rVert_F^2 \right)
\label{eq:hf_cc_loss}
\end{equation}
where $\mathbf{f}_{b,i}^{\rm ref,CC}$ are the reference forces computed at the CCSD(T)/cc-pVTZ level of theory, whereas the reference matrices $\mathbf{H}_b^{\rm core,ref}$ and $\mathbf{S}_b^{\rm ref}$ are taken from the HF/cc-pVDZ data. Note that there is a constant energy offset between CCSD(T)/cc-pVTZ and HF/cc-pVDZ data, which cannot be expressed when deriving energies with Eq.~\ref{eq:ground_state_en} without also modifying the core Hamiltonian, which is why CCSD(T)/cc-pVTZ level reference energies are not used in Eq.~\ref{eq:hf_cc_loss}. For any practical chemical application, only the relative energy between two structures is a well-defined quantity (absolute energies can be modified arbitrarily by adding constants without changing the underlying physics). To still be able to compute both energy and force errors with respect to the CCSD(T)/cc-pVTZ reference (see Section~\ref{Msec:results} in the main text), we analytically fit an energy shift constant that minimizes the squared error between energy predictions and reference energies in the training set and add it to the model prediction before evaluating it on the test set.

\bibliography{references}
\makeatletter\@input{xxms.tex}\makeatother